\documentclass[12pt,preprint]{aastex}

\def\Msun{M_\odot}
\def\gmajor{\left< \gamma^+ \right>}
\def\gminor{\left< \gamma^- \right>}

\def\go{\mathrel{\raise.3ex\hbox{$>$}\mkern-14mu
             \lower0.6ex\hbox{$\sim$}}}
\def\lo{\mathrel{\raise.3ex\hbox{$<$}\mkern-14mu
\lower0.6ex\hbox{$\sim$}}}

\slugcomment{revised manuscript, submitted to the Astrophysical Journal}

\begin{document}

\title{Multiple Weak Deflections and the Detection of Flattened Halos
with Galaxy--Galaxy Lensing}

\author{Candace Oaxaca Wright \& Tereasa G.\ Brainerd}

\affil{Boston University, Dept.\ of Astronomy, 725 Commonwealth Ave.,
Boston, MA 02215}

\begin{abstract}
We investigate the occurrence of multiple weak deflections in deep data sets
which are used to detect galaxy--galaxy lensing.  Using the
galaxies in the HDF (North) for which both redshifts and rest--frame blue
luminosities are known, we show that the probability for a given source
galaxy to be lensed by two or more foreground galaxies exceeds 50\% for
source redshifts $z_s \go 1$, and for which the
separate, individual deflections yield
$\gamma > 0.005$.  Neglecting multiple deflections when
obtaining best--fitting halo parameters for the lens galaxies, $\sigma_v^\ast$
and $s^\ast$, can lead to an overestimate by a factor of order 2 for
the characteristic halo mass interior to a radius of $100h^{-1}$~kpc.
We also show that multiple weak deflections
create systematic effects which may hinder
observational efforts to use weak lensing to constrain the projected
shapes of the dark matter halos of field galaxies.  We model the dark
matter halos of lens galaxies as truncated singular isothermal ellipsoids, and
for an observational data set in which the galaxies have
magnitudes in the range
$19 \lo I \lo 23$, we find that
multiple deflections result in strong correlations
between the post--lensing image shapes of most foreground--background
pairs of galaxies.  Imposing a simple redshift cut during 
the analysis of the data set, 
$z_d < 0.5$ and $z_s > 0.5$, is sufficient to reduce the correlation
between the final images of lenses and sources to the point that 
the expected anisotropy in the weak lensing signal can be detected via
a straightforward average.  We conclude that previous theoretical calculations
of weak lensing due to flattened dark matter
halos have considerably underestimated the
sizes of the observational
data sets which would be required to detect this effect.  In particular,
for a multi--color survey in which the galaxies have apparent magnitudes of
$19 \lo I \lo 23$ and the imaging quality is modest,
we find that a 4$\sigma$ detection should be obtained with a survey 
area of order 22~sq.~deg.,
provided 
photometric redshift estimates are made for the galaxies, the
typical error in $z_{\rm phot}$ is $\lo 0.1$, and only
source galaxies with azimuthal coordinates which are within $\pm 20^\circ$
of the lens symmetry axes are used in the data analysis.
\end{abstract}

\keywords{galaxies: halos -- dark matter -- gravitational lensing}

\section {Introduction}

The observed flatness of the rotation curves of the disks of spiral
galaxies, in addition to
the stellar dynamics and the hydrodynamics of hot gas in giant
ellipticals,
provides convincing evidence that most, if not all, large
galaxies reside within massive dark matter halos  
(see, e.g., Fabricant \& Gorenstein 1983; Stewart 
et al.\ 1984; Lauer 1985; and Fich \& Tremaine 1991 and references therein).
In addition, Zaritsky
\& White (1994) and Zaritsky et al.\ (1997) have used the dynamics of
genuine satellites of field spirals to investigate the gravitational potentials
of the halos, finding that the amount of dark matter associated with the
primary galaxy is large and, specifically, they conclude that
the mass contained within a
radius of $150 h^{-1}$~kpc of the primary galaxy is of order
1 -- 2$\times 10^{12} h^{-1} \Msun$ (here $h$ is the present value of the
Hubble parameter in units of 100 km/s/Mpc).  More recently, a dynamical
study 
of the masses of a subset of the Sloan Digital Sky Survey (SDSS) galaxies
has led to similar conclusions (McKay et al.\
2002).
Despite the apparent ubiquity
of dark halos matter
around galaxies, however, the characteristic physical parameters
associated with the halos, such as their shapes and physical extents,
remain poorly constrained.

A number of independent investigations have, however, shown 
that weak, systematic gravitational lensing of background galaxies by
foreground galaxies is a potentially powerful method by which the
physical parameters of galaxy halos may be constrained both in the
field and in rich clusters
(e.g., Brainerd, Blandford \& Smail 1996;
Dell'Antonio \& Tyson 1996; Griffiths et al.\ 1996; Hudson et al.\ 1998;
Ebbels 1998; Natarajan et al.\ 1998, 2001; Geiger \& Schneider 1999;
Fischer et al.\ 2000; Hoekstra 2000; Jaunsen 2000; McKay et al. 2001;
Smith et al. 2001;
Wilson et al.\ 2001).  
This effect, known as galaxy--galaxy lensing,
is too weak to be detected from the image of an individual 
lensed source, but since 
it acts coherently about the lens centers, it results in 
a slight preference for the images of distant source
galaxies to be oriented tangentially with respect to the 
locations of foreground lens galaxies on the sky. 
Unfortunately, it is difficult to compare all of the above studies
directly since the imaging quality,
image deconvolution methods, and the categorization of galaxies into
``lenses'' and ``sources'' is by no means consistent amongst the
investigations.  The diversity of both the data and 
the analysis techniques notwithstanding, however,
generally good agreement amongst the investigations has been found.  
In particular,
observations of galaxy--galaxy lensing by field galaxies have yielded 
inferred velocity dispersions for the halos of $L^\ast$ galaxies which
compare well with more traditional dynamical or hydrodynamical measurements
($\sigma_v \sim 140$~km/s to 190 km/s) and 
the inferred {\it maximum} radial extents of the halos of $L^\ast$ galaxies
are very large 
indeed ($\go 100h^{-1}$~kpc to $250h^{-1}$~kpc).

With the advent of very wide--field imaging capabilities, we
expect that galaxy--galaxy lensing will develop into a useful method
by which fundamental questions about galaxy formation can be addressed
directly.
For example, Natarajan et al.\ (1998, 2001) and Geiger \& Schneider (1999) 
have demonstrated that galaxy--galaxy lensing by the galaxies in rich
clusters can be used to constrain the degree to which the dark matter
halos of galaxies are truncated during infall.  Other potential uses
of galaxy--galaxy lensing include placing constraints on the evolution
of the total mass--to--light ratio of galaxies (both in the field and
in clusters), the morphological dependence of the halo potential (i.e.,
early--type versus late--type galaxies), the ``bias'' of light versus mass
in the universe (i.e., the galaxy--mass correlation function), and
even the shape of the redshift distribution of faint galaxies with 
redshifts in the range $1 \lo z \lo 3$ (see, e.g., Brainerd 2002; Brainerd
\& Blandford 2002).

In addition to these applications,
Natarajan \& Refregier (2000) and Brainerd \& Wright (2000) have 
discussed the possibility that observations of galaxy--galaxy lensing
could provide direct constraints on the projected shapes of dark
matter halos.  Although the simple singular
isothermal sphere can reproduce the flatness of the rotation
curves of the disks of spiral galaxies, there are both observational
and theoretical arguments in favor of halos which are flattened, rather
than spherical.  The observational evidence is somewhat scarce, owing to
the fact that there are relatively few galaxies for which the shape
of the halo potential can be probed directly
via traditional methods.
Nevertheless, the evidence for flattened halos is
quite diverse and includes such observations as
the dynamics of polar ring galaxies, the
geometry of X-ray isophotes, the flaring of HI gas in spirals, the
evolution of gaseous warps, and the kinematics of Population II stars
in our own Galaxy.  In particular, studies of disk systems which probe
distances of order 15~kpc from the galactic planes suggest that the
ratio of shortest to longest principle axes of the halos is
$c/a = 0.5 \pm 0.2$ (see, e.g., the comprehensive review by Sackett
1999 and references therein). 
Studies of a number of strong lens galaxies have also suggested that the
mass distributions of the lenses are not precisely spherical.
For example, Maller et al.\ (2000) have 
found that, provided the disk mass is small compared to the halo mass,
the halo of the spiral galaxy which lenses the quasar B1600+434 is
consistent with $c/a = 0.53$.  
In addition, the 17 strong lens systems studied by Keeton, Kochanek \&
Falco (1998) showed some preference for flattened mass distributions, although
extremely flattened (i.e., ``disky'') mass distributions were ruled out.
Finally, high--resolution simulations
of dissipationless cold dark matter models consistently result in 
markedly non--spherical  galaxy halos 
with a mean projected ellipticity of
order 0.3 (e.g., Dubinski \& Carlberg 1991; Warren et al.\ 1992)
and, therefore, from a theoretical standpoint it is
not at all unreasonable to expect that the dark matter halos of galaxies
should be somewhat flattened in projection.

Unlike a spherically--symmetric lens for which the gravitational lensing
shear is isotropic about the lens center, the shear due to an elliptical
lens is anisotropic about the lens center.  
Specifically, at a given angular distance, $\theta$,
from an elliptical lens, 
source galaxies which are located closer to the major axis
of the mass distribution of the lens
will experience greater shear than sources which
are located closer to the minor axis (e.g., Schneider, Ehlers \& Falco
1992).  Noting this well--known effect, 
Natarajan \& Refregier (2000) and Brainerd \& Wright (2000) modeled
the dark matter halos of field galaxies as infinite
singular isothermal ellipsoids
and made rough estimates of the sizes of observational data sets which would
be required to detect ``anisotropic'' galaxy--galaxy lensing and, hence,
to constrain the net flattening of the halo population.  Both studies conclude
that, if the mean flattening of the halos is of order 0.3, then forthcoming
wide--field surveys such as the SDSS should be able to
detect this effect.

In estimating the amount of data which would be required to detect anisotropic
galaxy--galaxy lensing, 
both Natarajan \& Refregier (2000) and Brainerd \& Wright (2000) made the
simplifying assumption that each distant source galaxy is lensed by only
one foreground galaxy.  However, for a modestly deep imaging survey 
($I_{\rm lim} \sim 23$), the simulations of galaxy--galaxy
lensing performed by Brainerd, Blandford
\& Smail (1996), hereafter BBS, indicated that most of the galaxies with
magnitudes in the range
$22 \lo I \lo 23$ would, in fact,
 have been lensed at a comparable level by two or
more foreground galaxies (see, e.g., \S3.5 of BBS).  
Therefore we expect that in a realistic data set,
such ``multiple deflections'' may significantly affect the signal--to--noise
which could be achieved when attempting to detect anisotropic galaxy--galaxy
lensing.  

Although BBS included a largely qualitative discussion of the importance
of multiple weak deflections in deep data sets, a detailed quantitative
discussion of this effect has not been published to date.  Therefore, we
begin this paper by demonstrating that multiple weak deflections are
expected to occur with high probability within sufficiently deep data sets.
For this we use the well--studied northern Hubble Deep Field, hereafter
HDF--N, (Williams et al.\ 1996).  In \S2
we compute the probability distribution of weak galaxy--galaxy
lensing events for the 
faint galaxies in the HDF--N as a function of the number of deflectors
(i.e., the number of foreground lenses).  In addition,
we briefly investigate the effect of
ignoring such multiple deflections on the values of the
dark matter halo parameters
that are inferred from observations of galaxy--galaxy lensing.

Having established the importance of multiple weak deflections
in deep data sets, we then
investigate the systematic lensing of background
galaxies by foreground galaxies for a case
in which the halos of the lens galaxies are
modeled as truncated, singular isothermal ellipsoids.  We perform 
Monte Carlo simulations of galaxy--galaxy lensing, 
including the effects of multiple weak deflections
on the final images of distant galaxies, and we compute the
area of a deep ground--based
imaging survey ($I_{\rm lim} \sim 23$) which would be required
in order to detect anisotropic galaxy--galaxy lensing via a straightforward
average of the signal.  The details of
the simulations, including a discussion of the lensing
properties of the dark matter halos, are given in \S3.
The anticipated anisotropy in the galaxy--galaxy lensing signal and the
results for the area of a survey which would be necessary in order 
to detect anisotropic galaxy--galaxy lensing
in the presence of realistic observational noise are given in \S4.
A discussion of our results is presented in \S5.

\section{Multiple Weak Deflections in the HDF--N}

The Hubble Deep Field (North) and flanking fields have
been the subject of a deep redshift
survey (Cohen et al.\ 2000) as well as an extensive multicolor photometric
investigation (Hogg et al.\ 2000).  As a result, both the redshifts, $z$,
and the rest--frame blue luminosities, $L_B$, of $\sim 600$ galaxies in this
region of space are known (Cohen 2002).  Therefore, it is possible to make
quite a detailed theoretical prediction for the weak galaxy--galaxy lensing
shear field in the region of the HDF--N and, specifically, 
for the probability of 
multiple deflections.

For simplicity, we follow the approach of BBS in their galaxy--galaxy
lensing simulations and we scale the physical properties of the dark
matter halos around the galaxies in the HDF--N and the flanking fields
in terms of the characteristic
properties associated with the halos of $L^\ast$ galaxies.  
We let $\sigma_v^\ast$ be the velocity dispersion of an $L^\ast$ galaxy halo
and we assume that a Tully--Fisher or Faber--Jackson type of relation will hold
for each of our galaxies. We therefore have
\begin{equation}
\frac{\sigma_v}{\sigma_v^\ast} = \left( \frac{L_B}{L_B^\ast} \right)^{1/4},
\end{equation}
where $\sigma_v$ is the velocity dispersion of a halo in which a galaxy with
luminosity $L_B$ resides.  

In this section we will focus on the probability of multiple deflections and
we will compute only the azimuthally--averaged galaxy--galaxy lensing shear
(i.e., the mean shear within circular annuli that are centered on the lens
galaxies).  In this case it is sufficient to adopt a circularly--symmetric
density profile for the dark matter halos and, again, we adopt the 
density profile used by BBS in their analysis:
\begin{equation}
\rho(r) = \frac{\sigma_v^2 s^s}{2\pi G r^2 \left( r^2 + s^2 \right) },
\end{equation}
where $G$ is Newton's constant and $s$ is a characteristic halo radius.
(Note that since our analysis focuses solely on the weak lensing regime, it
is sufficient to adopt a density profile with a central density which is 
singular since the inclusion of any reasonable--sized core will not affect
the weak lensing properties of the halos at all.)
As in BBS we assume that the total
mass--to--light ratio of a galaxy is constant independent of its luminosity
and, therefore, the radii of the halos of galaxies with $L_B \ne L_B^\ast$
scale with the radii of the halos of
$L_B^\ast$ galaxies according to:
\begin{equation}
\frac{s}{s^\ast} = \left( \frac{L_B}{L_B^\ast} \right)^{1/2}.
\end{equation}
Throughout we adopt fiducial values of $\sigma_v^\ast = 156$~km/s
and $s^\ast = 100h^{-1}$~kpc for the halos of $L^\ast_B$ galaxies
(i.e., comparable to the characteristic parameters which have been
inferred from previous studies of galaxy--galaxy lensing).

Having made these assumptions, we can make predictions for
the galaxy--galaxy lensing shear field within the region of the
HDF--N that would 
be generated by the 
$\sim 600$ galaxies in Cohen (2002).  These objects include not
only galaxies in the HDF--N, but also galaxies in the flanking fields and,
due to the substantial masses
of some of the galaxies in the flanking fields, their weak lensing effects
may influence the shear field inside the much smaller region of
the HDF--N itself.  We
therefore include the flanking field galaxies
in all of our calculations for the weak shear field inside the HDF--N.

Here our goal is to quantify the probability that a given faint galaxy
will be weakly--lensed by one or more foreground galaxies.  
This probability will, of course, be a strong
function of the actual value of the shear, $\gamma$, due to a given
weak lensing deflection.  That is, it is
much more likely for a distant galaxy to be lensed by a foreground
galaxy which
produces a shear of 
$\gamma \sim 0.001$ than, say, a
shear of $\gamma \sim 0.01$.  Therefore, in order to discuss the number
of weak deflections that a given source galaxy is likely to encounter, we
must first decide what minimum value of $\gamma$ qualifies as a
``significant'' deflection.

A typical value for the net shear due to galaxy--galaxy lensing is
$\gamma \sim 0.005$ (see, e.g., the observational papers cited in the
Introduction) and we use this value of $\gamma$ as a baseline
for computing the number of weak lensing deflections that source galaxies
undergo.  Specifically, we compute $P(N_D)$, the probability
that a given galaxy is lensed by $N_D$ foreground galaxies (i.e, 
``deflectors'') where each {\it individual deflection} gives rise to 
a shear of $\gamma \ge 0.005$.  That is, $P(N_D = 2)$ is the probability that
a given galaxy has been lensed by two individual
foreground galaxies, each of which
lensed the distant galaxy at a level which is comparable to or greater
than the expected net shear due to galaxy--galaxy lensing.

\begin{figure}
\plotone{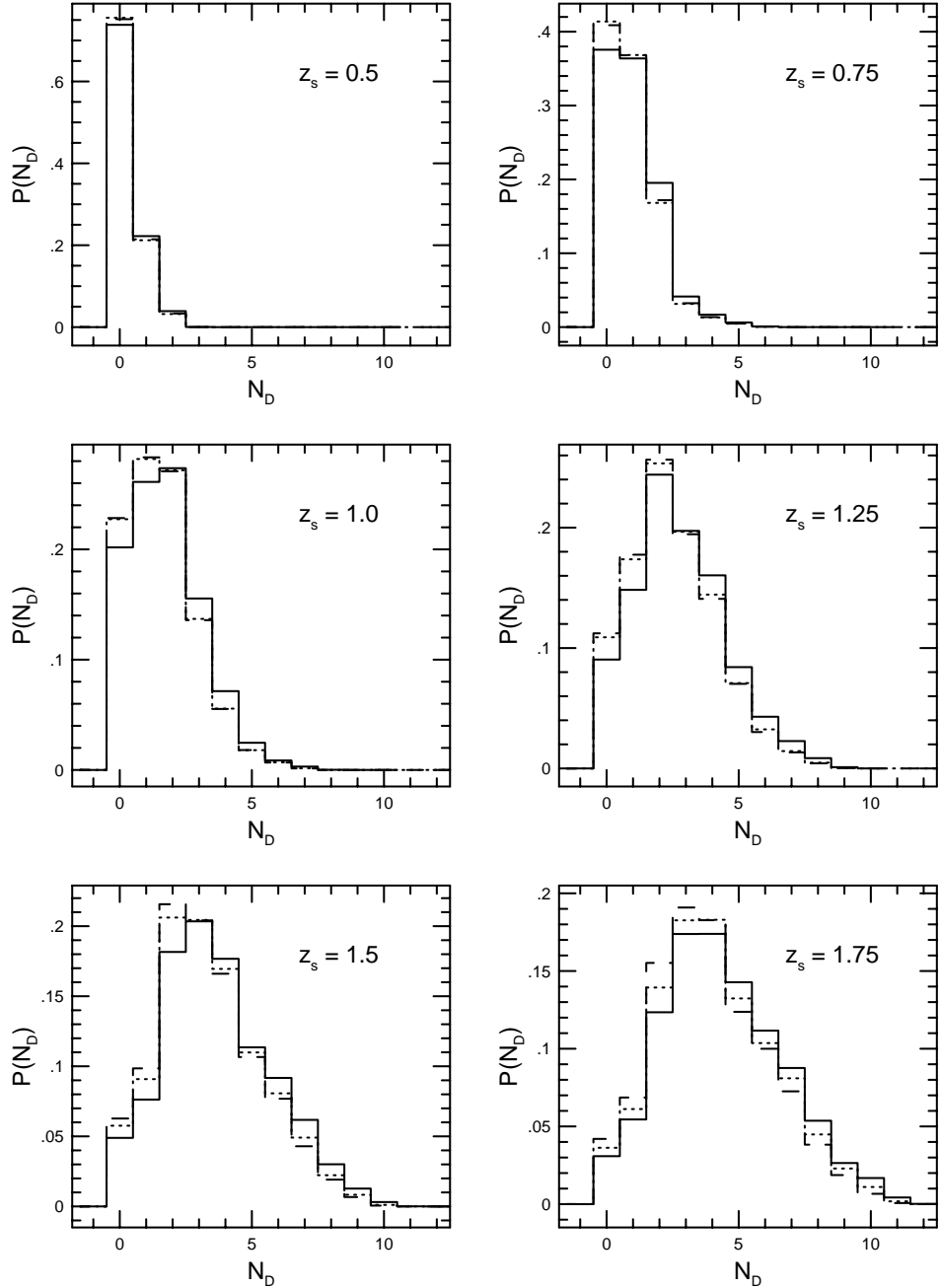}
\vskip -0.5in
\caption{Probability distribution of weak lensing deflections
for a source galaxy with redshift $z_s$ that is located
in the HDF--N.  Only individual
deflections
which give rise to $\gamma > 0.005$ are included in $P(N_D)$.  
For sources with $z_s > 1$, the probability of multiple deflections
(i.e., $N_D \ge 2$) exceeds 50\%.  Results
are shown for three cosmologies: open (solid line), flat Lambda--dominated
(dotted line), and Einstein--de Sitter (dashed line).
\label{pdist}}
\end{figure}

Shown in Fig.\ \ref{pdist} is the theoretical probability distribution
function, $P(N_D)$, for source galaxies in the HDF--N with redshifts
of $z_s$.   In order to produce this result, source galaxies were
placed at random locations within the HDF--N and were assigned specific
redshifts, $z_s$.  
Note that since the angular clustering of cosmologically--distant galaxies
is observed to be quite weak (e.g., Infante \& Pritchett 1995; Villumsen,
Freudling \& da Costa 1997; Brainerd \& Smail 1998) and since BBS found
that the inclusion of the intrinsic clustering of the faint galaxies in their
sample made no measurable difference to the predicted galaxy--galaxy lensing 
signal, the use of random location for our source
galaxies is reasonably well--justified.

Using the scaling relations above, the individual
values of $\gamma$ due to each one of the galaxies in Cohen (2002) is
computed for the source galaxies and a probability distribution for
individual deflections with $\gamma > 0.005$ is assembled.
Results are shown in Fig.\ \ref{pdist} for three different cosmologies:
open ($\Omega_0 = 0.3$, $\Lambda_0 = 0$, $H_0 = 60$~km/s/Mpc),
flat Lambda-dominated ($\Omega_0 = 0.3$, $\Lambda_0 = 0.7$,
$H_0 = 60$~km/s/Mpc), and Einstein--de Sitter  ($\Omega_0 = 1.0$, 
$\Lambda_0 = 0$, $H_0 = 60$~km/s/Mpc).  The probability 
distributions
are nearly independent of the cosmology, which is unsurprising since
the galaxy--galaxy lensing shear has only a weak cosmology dependence
(see, e.g., BBS \S3.6).  

For a source galaxy in the HDF--N which is located at a redshift of $z_s = 0.5$,
the probability of being lensed
by even one foreground galaxy is quite small.
This is due to the fact that the median redshift of
the galaxies in Cohen (2002) is of order 0.6 and, hence, many of the
``lens'' galaxies are actually more distant than the
``source'' galaxies.
At a redshift of $z_s = 1$, however, the probability that a
source galaxy has been lensed at a significant level by {\it two or more}
foreground galaxies is of order 50\%.  At source redshifts of
$z_s = 1.25$, 1.5, and 1.75, the probability of a source galaxy encountering
multiple deflections of $\gamma > 0.005$ increases to of order
70\%, 85\%, and 90\%, respectively.  Therefore, we expect that in a 
deep data set for which
the median redshift is $\go 1$, multiple weak deflections of a
substantial magnitude are very likely to have occurred. Of course, in the
case of individual deflections for which $\gamma < 0.005$, the probability
of comparable multiple deflections occurring at any given source redshift
will be greater than the results shown in Fig.\ \ref{pdist}.

A major goal of studies of galaxy--galaxy lensing is to
constrain the nature of the dark matter halos of the lens galaxies.  That 
is, one wishes to use observations of the galaxy--galaxy lensing shear
to measure the values of the parameters $\sigma_v^\ast$ and $s^\ast$ above.
However, it is clear from Fig.\ \ref{pdist} that multiple deflections 
will likely occur in deep data sets and, therefore, they should be included
in calculations which attempt to infer $\sigma_v^\ast$ and $s^\ast$
from a detection of galaxy--galaxy lensing.  
To assess the degree of error in $\sigma_v^\ast$ and $s^\ast$ which 
would be caused by not including multiple deflections, we
compute $\gamma(\theta)$,
the theoretical mean shear as a function of lens--source separation in
the HDF--N, for two distinct cases: (i) all of the weak deflections
are included in the determination of the net shear experienced by
each source galaxy, and (ii)
only the deflection due to the closest lens (in projection on the sky) is
used to determine the shear.  That is, in case (ii) we 
ignore all of the multiple deflections and we
assume that the nearest neighbor
lens galaxy is the only lens.  This is motivated by the fact that for
a singular isothermal potential the shear decreases as $\gamma(\theta) \propto
\theta^{-1}$ and, 
therefore, in the simplistic case of lenses being confined to a
single plane in redshift space and sources being confined to another plane in 
redshift space, the nearest neighbor lens is likely to be the only ``important''
lens.

The magnitude of the mean shear due to galaxy--galaxy lensing is, of
course, strongly--dependent upon the depth of the source sample and
here we restrict our analysis to source galaxies with
$I < 23$ (corresponding to a median redshift of order
0.85).  Although the completeness limit of the
HDF--N extends far below this magnitude,
$I_{\rm lim} \sim 23$ is comparable to the limiting magnitude of
most of the deeper data sets which
have so far been used to study galaxy--galaxy lensing. Unfortunately,
due to the small area of the field there are relatively few
galaxies in the HDF--N data which satisfy our limiting magnitude
requirements (only of order 30 galaxies with known redshifts).  Therefore, an
accurate prediction of the effects of multiple
deflections on the galaxy--galaxy lensing shear cannot be obtained
simply by using the observed HDF--N galaxies.  Instead, we must generate
Monte Carlo source galaxy catalogs which reproduce the observed faint galaxy
number counts and for which we adopt a parameterized redshift distribution.
As in BBS, we use a redshift distribution of the form
\begin{equation}
P(z|I) = \frac{\beta z^{2}\exp[-(z/z_{0})^{\beta}]}{\Gamma(3/\beta)z_{0}^{3}}
\label{zdist}
\end{equation}
(e.g., Baugh \& Efstathiou 1993),
which is in good agreement with both the no--evolution model and with the 
observed redshift distribution of galaxies with $ 19 \le I \le 22$ 
obtained from
the Canada France Redshift Survey
(e.g., LeF\`evre et al.\ 1996).  Here
\begin{equation}
z_{0} = k_{z}[z_{m} + z_{m}'(I - I_{m})],
\end{equation}
where $z_m$ is the median redshift, $I_m$ is the median $I$-band
magnitude, and $z_m'$ is the derivative of the median redshift with respect
to $I$.  Extrapolating the results of the CFRS to a sample of galaxies with 
$19 \le I \le 25$ we have that $z_m = 0.86$, $z_m' = 0.15$, $k_z = 0.8$,
and $\beta = 1.5$.  A total of 6250 Monte Carlo realizations of the 
source galaxy distribution were generated, and the shear experienced
by each of the sources was computed.

Since
galaxy--galaxy lensing occurs in the weak regime (i.e., $| \vec{\gamma} |
<< 1$), the net shear experienced by any source galaxy which has been 
weakly--lensed by more than one foreground galaxy can be computed simply
as the (complex) vector sum of the shears induced by each of the individual
lenses (e.g., Schneider, Ehlers \& Falco 1992).  Therefore, the net shear
experienced by a given source galaxy after having been lensed by
$M$ weak lenses is given by:
\begin{equation}
\vec{\gamma} \equiv \gamma_1 + i \gamma_2 = \sum_{j=1}^M \gamma_{1,j} 
+ i \sum_{j=1}^M \gamma_{2,j},
\end{equation}
where $\gamma_{1,j}$ is the real component of the shear due to lens
$j$ and $\gamma_{2,j}$ is the imaginary component.
For our calculations of the net shear due to multiple weak deflections,
we include all deflections, $j$, encountered by each source galaxy (i.e., not
just those deflections for which $\gamma_j > 0.005$).

\begin{figure}
\plotone{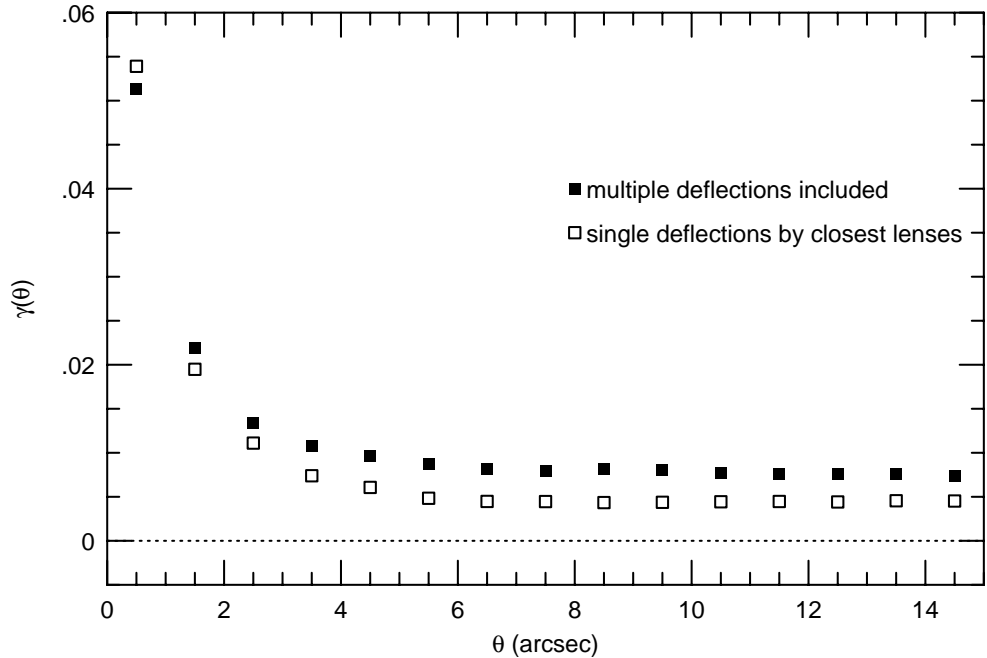}
\vskip -3.0in
\caption{Circularly--averaged shear for galaxies in the HDF--N. An
open cosmology with $\Omega_0 = 0.3$, $\Lambda_0 = 0$,
and $H_0 = 60$~km/s/Mpc has been adopted for this figure, but the results
are essentially independent of the choice of cosmology.
Solid squares: multiple deflections are used in the determination of
the net shear for a given source galaxy.  Open squares: only the shear
due to the nearest--neighbor lens is used to determine the shear for
a given source galaxy.  Source galaxies with $19 < I < 23$ and a 
magnitude--redshift relation obtained from an extrapolation of the 
CFRS redshift
distribution are used in the calculation (see text).
\label{gtheta}}
\end{figure}

\begin{figure}
\vskip -1.0in
\plotone{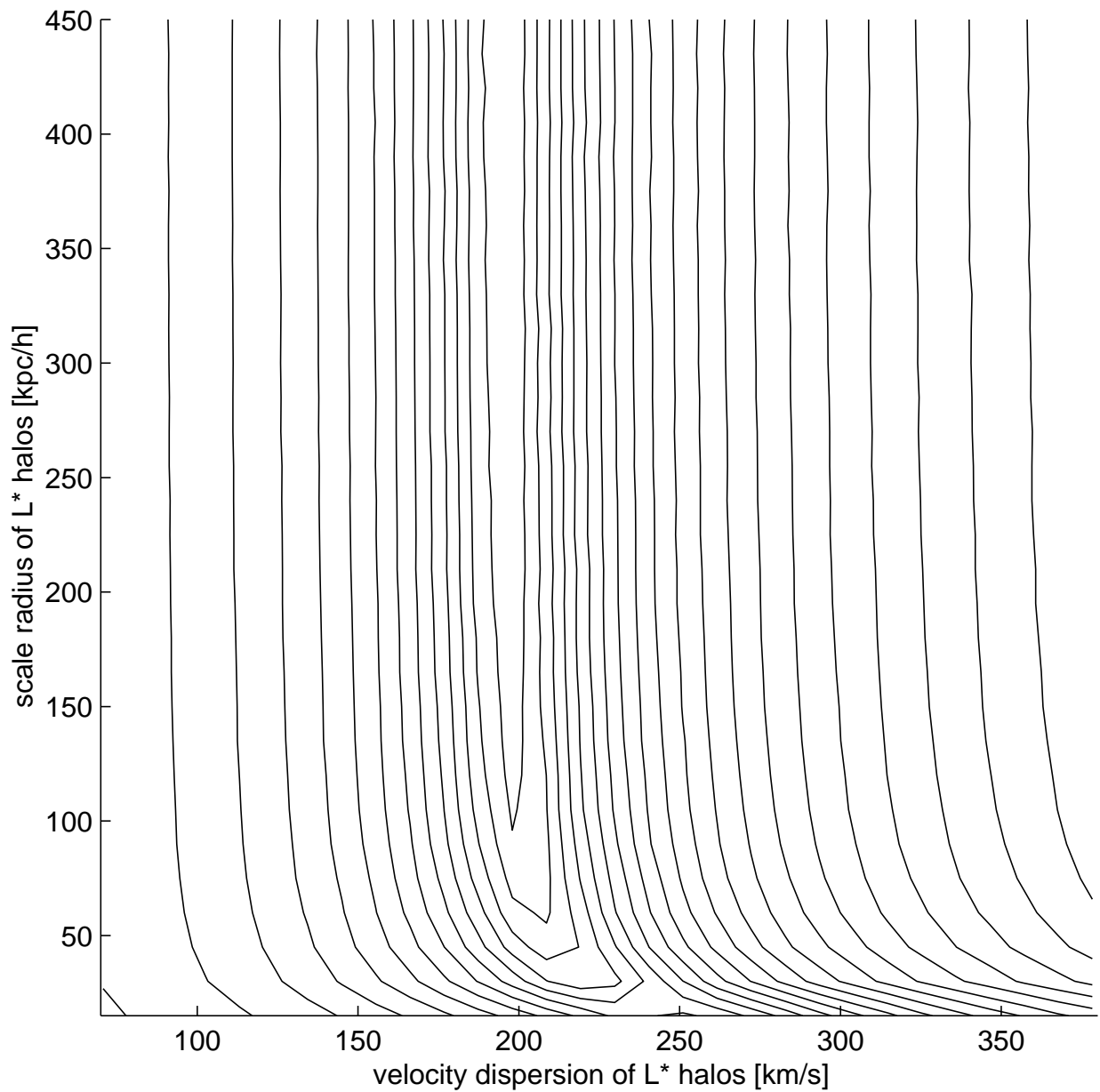}
\caption{Logarithmically--spaced contours of constant $\chi^2$ for
a comparison of
$\gamma(\theta)$ obtained from single--deflection
calculations (using various values of $\sigma_v^\ast$ and $s^\ast$) with
$\gamma(\theta)$ obtained from
a multiple--deflection calculation in which $\sigma_v^\ast = 156$~km/s and
$s^\ast = 100h^{-1}$~kpc.  
\label{chi2}}
\end{figure}

Shown in Fig.\ \ref{gtheta} is the variation of the mean shear with 
lens--source separation, both with and without the inclusion of multiple
deflections.  As before, we take $\sigma_v^\ast = 156$~km/s and
$s^\ast = 100h^{-1}$~kpc and, for simplicity, we have only considered
the case of an open universe.  (Since the galaxy--galaxy
lensing signal is so weakly--dependent upon the cosmology, there is
no significant loss of generality from this choice.)
On scales $> 1''$, Fig.\ \ref{gtheta} shows that there is a substantial
systematic error (of order 50\%) in the mean shear when
multiple deflections are ignored in the calculation.  As a result, if 
one were to model the true shear field with a theoretical shear field
that did not include multiple deflections, substantial systematic errors
in the best--fitting values of $\sigma_v^\ast$ and $s^\ast$
would result.  We demonstrate this in Fig.\ \ref{chi2}, where
we compare the actual shear produced by the HDF--N galaxies (i.e., the
solid squares in Fig.\ \ref{gtheta}) to the shear that one obtains for
single--deflection calculations which use various values of $\sigma_v^\ast$ 
and $s^\ast$.  Formally, the best--fitting values from the single--deflection
calculations are $\sigma_v^\ast = 198$~km/s and $s^\ast = 435h^{-1}$~kpc,
which are
very much larger than the input values of 156~km/s and 100$h^{-1}$~kpc.
(Note, however, that the galaxy--galaxy lensing shear is relatively insensitive
to the radii of the halos and, so, for a given value of $s^\ast$,
$\chi^2$ is nearly identical for all values of
$s^\ast$ which exceed $100h^{-1}$ kpc.  See also \S3.6 of BBS.)
Neglecting multiple deflections in the process of constraining $\sigma_v^\ast$
and $s^\ast$ leads, therefore, to a formal estimate of the mass of
the halo of an $L^\ast$ galaxy which is a factor of $\sim 2$ too large
within a radius of
$100h^{-1}$~kpc (e.g., BBS equation 3.4).

We therefore conclude that the effects of multiple weak deflections can
be quite important in the accurate prediction of theoretical galaxy--galaxy
lensing shear fields, and in the determination of the characteristic 
parameters which are associated with the halos of the lens galaxies.  We
now turn to theoretical calculations of shear fields caused by flattened
dark matter halos, and for which we have explicitly included all of the
multiple deflections.

\section{Galaxy--Galaxy Lensing with Flattened Halos}

While the HDF--N is a useful field for demonstrating the likelihood
and effects of multiple weak deflections, it is nevertheless an extremely
small field and one for which the effects of halo flattening will not
be convincingly detected in the galaxy--galaxy lensing signal.  We, therefore,
have constructed large Monte Carlo galaxy catalogs in order to 
investigate weak lensing by non--spherical galaxy halos.
In the following subsections we outline the details of the galaxy catalogs
and their weak lensing properties.

\subsection{Luminous Properties of the Galaxies}

The Monte Carlo galaxy catalogs are designed to reproduce a number of
observational constraints on the faint galaxy population:
\begin{itemize}
\item the number counts of faint galaxies, $\frac{d\log N}{dm}$, to a
limiting magnitude of $I_{\rm lim} = 25$
\item the shape of the redshift distribution of faint galaxies, $N(z)$,
extrapolated to $I_{\rm lim} = 25$
\item the distribution of intrinsic
shapes (i.e., the unlensed equivalent image ellipses of the light 
distribution of the sources),
as obtained from deep imaging with HST
\end{itemize}
Each galaxy catalog is generated within a $40' \times 40'$ region of
sky and for simplicity we adopt
an open cosmology with $H_0 = 60$~km/s/Mpc, $\Omega_0 = 0.3$,
$\Lambda_0 = 0$. A total of 500 independent catalogs were
generated and were
then used to compute the mean galaxy--galaxy lensing signal which
would occur if the dark matter halos of the galaxies were substantially
flattened.

As in \S2, all galaxies in our catalogs
are assigned random locations on the frame.  Each galaxy is also
assigned an apparent magnitude, $19 < I < 25$, which is drawn from the deep
$I$--band number counts obtained by Smail et al.\ (1995). This results
in a number density of 60 galaxies per square arcminute in the simulations.  In
addition, each galaxy is assigned an intrinsic (i.e., unlensed)
position angle which is
drawn at random from $- \frac{\pi}{2} \le \phi \le \frac{\pi}{2}$.  That is,
we assume that in the absence of gravitational lensing, the orientations of
the major axes of
the images of the distant galaxies are randomly distributed.
The galaxies are also assigned intrinsic shapes,  
$\tau \equiv (a^{2} - b^{2})/(2ab)$,
drawn from the probability distribution
\begin{equation}
P(\tau) = A\tau\exp[-(\tau/0.036)^{0.54}]
\label{taudist}
\end{equation}
which was obtained by Ebbels (1998) using 94 deep
archival HST images of blank fields.
Here $a$ is the major
axis of the (unlensed) 
equivalent image ellipse of a galaxy and $b$ is the minor axis.

Each galaxy is assigned a redshift based upon its
apparent magnitude, drawn from equation (\ref{zdist}) in \S2, and
each galaxy is also assigned an intrinsic luminosity, $L$, according to:
\begin{equation}
\frac{L}{L^\ast} = \left(\frac{H_0 D_{d}}{c}\right)(1+z)^{3+\alpha}
10^{0.4(22.9-I)},
\end{equation}
(see, e.g., BBS). Here 
$D_d$ is the angular diameter distance to the
galaxy, $z$ is its redshift, $I$ is its apparent magnitude, and the
parameter $\alpha$ is the slope of the spectral energy distribution:
\begin{equation}
\alpha = - \frac{d\log_{10} L_{\nu}}{d\nu}.
\end{equation}
For simplicity we adopt the mean slope, $\alpha = 0.42$, between the
$R$ and $B$ bands which was found in the Caltech Faint Galaxy Redshift Survey
(Cohen et al.\ 1999ab).

\subsection{Gravitational Properties of the Galaxies}

Having assigned all of the intrinsic properties of the galaxies which are
associated with the light distribution, 
we then assign the intrinsic properties
which are associated with the distribution of the
mass.  Each galaxy is assumed to reside within
a dark matter halo which can be fairly represented as a truncated singular
isothermal ellipsoid.  Following Kormann et al.\ (1994), the surface mass
densities of the dark matter halos are given by
\begin{equation}
\Sigma(\rho)  = \frac{\sigma_v^{2}\sqrt{f}}{2G}
\left(\frac{1}{\rho} - \frac{1}{\sqrt{\rho^{2} + x_{t}^{2}}}\right),
\label{surfmass}
\end{equation}
where $\sigma_v$ is the line of sight velocity dispersion, 
$f$ is the
axis ratio of the 
mass distribution as projected on the sky ($0 < f \le 1$), $x_t$ is a truncation radius, $G$ is Newton's constant,
and $\rho$ is a generalized elliptical
radius defined such that $\rho^2 = x_1^2 + f^2 x_2^2$.  Here
$x_1$ and $x_2$ are
Cartesian coordinates measured, respectively, along the minor and major
axes of the projected mass distribution of the halo.  In the limiting case of
a round lens (i.e., $f \rightarrow 1$), the total mass of the halo becomes
\begin{equation}
M_{\rm tot} = \frac{\pi \sigma_v^2 x_t}{G},
\end{equation}
which is identical to the halo mass model adopted by BBS.

\begin{figure}
\plotone{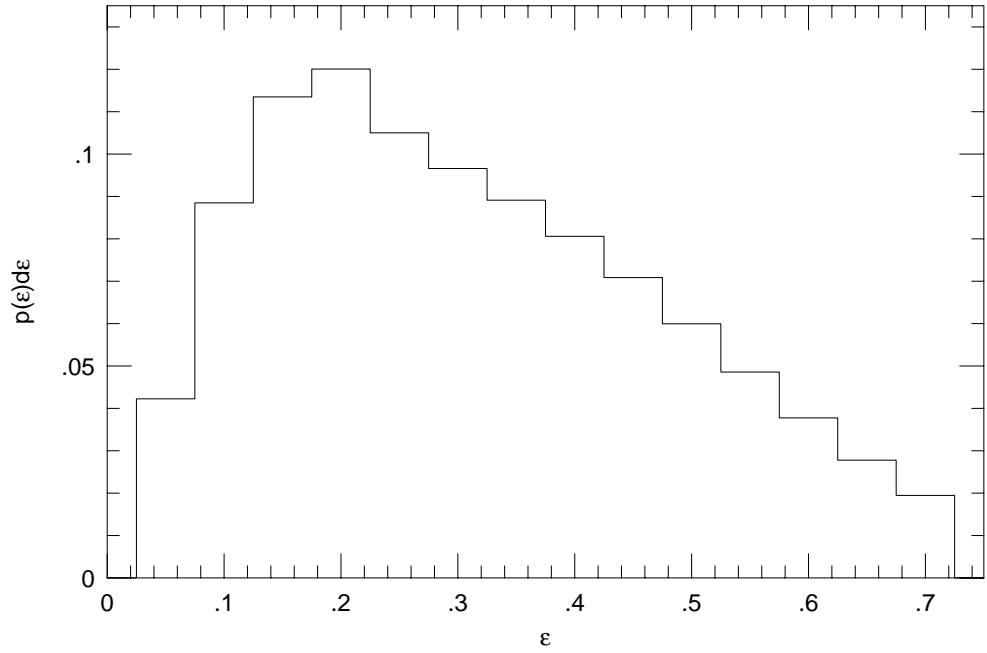}
\vskip -4.0in
\caption{The distribution of projected ellipticities of the
halos.  Here $\epsilon \equiv 1 - b/a = 1 - f$. \label{edist}}
\end{figure}

Provided a galaxy is in a state of dynamical equilibrium (i.e., it has
not undergone a recent collision or merger), it is reasonable to expect that,
in projection on the sky, the position angle of major axis of the mass
distribution will be aligned well with the (unlensed)
major axis of the light distribution.  In addition, work by
Keeton, Kochanek \& Falco  (1998) and Kochanek (2001) has shown that
the mass and light in strong lens galaxies is aligned quite well.
Therefore, we assign a position angle to each halo (as projected on the
sky) which is identical to the intrinsic position angle that was assigned
to its light distribution in \S3.1.  However, we do not assign identical
intrinsic ellipticities to both the mass and light
distributions.  Instead,
the ellipticity of each galaxy's halo is drawn from a probability distribution
which is constructed from current observational constraints
on the shapes of galaxy halos.  Specifically, from principle moment analyses
in which, by definition, $a > b > c$, 
the distribution of halo shapes seems to favor
$c/a = 0.5 \pm 0.2$ and $b/a \go 0.8 $ (see, e.g.,
Sackett 1999).  To generate
the distribution of projected ellipticities of the halos, then, we construct
ten million ellipsoids in which the values of
$b/a$ are drawn from a uniform distribution
$0.8 \le b/a \le 1.0$ and the values of
$c/a$ are drawn from a Gaussian distribution with a
mean of 0.5 and a standard deviation of 0.2.  Each ellipsoid is then projected
onto a plane which is oriented perpendicularly to a randomly--chosen line
of sight and the distribution of the projected axis ratios of the ellipsoids
is computed.  The result is shown in Fig.\ \ref{edist}, 
where the mean and median
ellipticities of the distribution are both of order 0.3 and we
use the convention that $\epsilon \equiv 1 - b/a = 1 - f$.

As in \S2, we scale the physical properties of the galaxy halos in terms
of the characteristic properties associated with the halos of $L^\ast$ 
galaxies:   
$\frac{\sigma_v}{\sigma_v^\ast} = \left( \frac{L}{L^\ast} \right)^{1/4}$ and
$\frac{x_t}{x_t^\ast} = \left( \frac{L}{L^\ast} \right)^{1/2}$.
Again, we adopt a fiducial value of
$\sigma_v^\ast = 156$~km/s for the velocity dispersion
of the halo of an $L^\ast$ galaxy 
and a corresponding truncation
radius of $x_t^\ast = 100h^{-1}$~kpc.

\subsection{Lensing Properties of the Halos}

The convergence of the lenses is given by
\begin{equation}
\kappa(\rho) = 
\frac{\sigma_v^{2}\sqrt{f}}{2G \Sigma_c}
\left(\frac{1}{\rho} - \frac{1}{\sqrt{\rho^{2} + x_{t}^{2}}}\right),
\label{kappaeq}
\end{equation}
where $\Sigma_c \equiv  \left(\frac{4\pi G}{c^{2}}
\frac{D_{d}D_{ds}}{D_{s}}\right)^{-1}$ is the ``critical surface mass
density'', $D_d$ is the angular diameter
distance of the lens, $D_s$ is the angular diameter distance of the source
and $D_{ds}$ is the angular diameter distance between the lens and the source.
The real and imaginary 
components of the shear due to a lens with a projected axis ratio of $f$
can be derived straightforwardly by using equations
(63abc) of Kormann et al.\ (1994):
\begin{eqnarray}
\gamma_{1} & = & \frac{\sigma_v^{2}\sqrt{f}}{2G\Sigma_{c}}
\left[-\frac{\cos(2\varphi)}{\rho} - 
\left\{f^{2}\left(x_{1}^{2}-x_{2}^{2}\right) - \left(1-f^{2}\right)x_{t}^{2}
\right\}{\mathcal P}\right] \\
\gamma_{2} & = & \frac{\sigma_v^{2}\sqrt{f}}{2G\Sigma_{c}}
\left[-\frac{\sin(2\varphi)}{\rho} - 
2f^{2}x_{1}x_{2}{\mathcal P}\right]
\end{eqnarray}
where
\begin{equation}
{\mathcal P} \equiv \frac
{x_{1}^{2}+f^{4}x_{2}^{2}-
(1+f^{2})(\rho^{2}+x_{t}^{2})+
2fx_{t}\sqrt{\rho^{2}+x_{t}^{2}}}
{\sqrt{\rho^{2} + x_{t}^{2}}
\left[f^{4}r^{4}-
2f^{2}(1-f^{2})x_{t}^{2}(x_{1}^{2}-x_{2}^{2})+
(1-f^{2})^{2}x_{t}^{4}\right]}.
\label{peq}
\end{equation}
Recall from eqn.\ (\ref{surfmass}) 
above that $x_1$ and $x_2$ are Cartesian coordinates
measured along the minor and major axes of the lens, respectively.
In order to maintain consistency with the notation of Kormann et al.\ (1994),
here we have also introduced a polar coordinate system, centered on the lens,
with radial coordinate $r \equiv \sqrt{x_1^2 + x_2^2}$ and polar angle,
$\varphi$, defined  such that $x_1 = r \cos \varphi$ and $x_2 = r\sin \varphi$.

\begin{figure}
\plotone{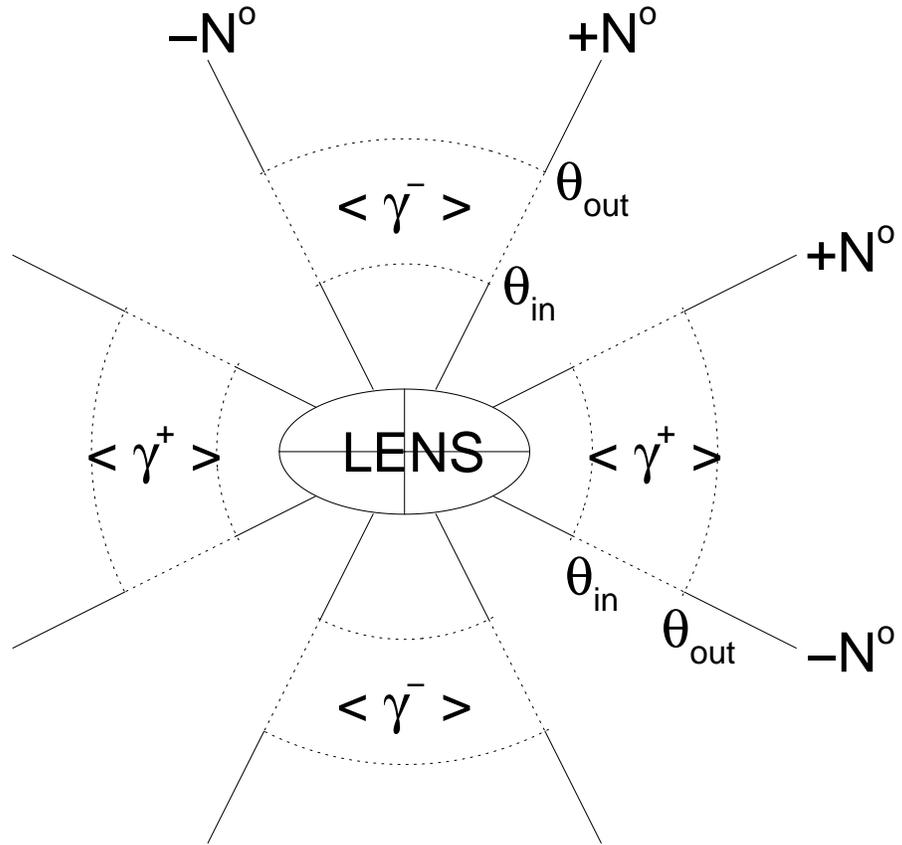}
\vskip -2.0in
\caption{Illustration of our notation.  The shear experienced
by sources located within $\pm N^\circ$ of the major axis of
the lens ($N \le 45^\circ$)
is denoted by $\gamma^+$ and, similarly, 
the shear experienced by sources located
within $\pm N^\circ$ of the minor axis ($N \le 45^\circ$) is 
denoted by $\gamma^-$.
Only sources which are 
located within the dashed regions (defined by the value of
 $N$ and by the sizes of the 
inner and outer angular radii, $\theta_{\rm in}$ and $\theta_{\rm out}$)
are used in the calculations of the mean values of the shear,
$\left<\gamma^+ \right>$ and
$\left<\gamma^- \right>$. \label{diagram}}
\end{figure}

It is clear from equations (\ref{kappaeq}) through (\ref{peq}) that, 
unlike
a circularly symmetric lens for which the magnitude of the shear depends
only upon the angular distance from the lens center, the shear due to an
elliptical lens is a function of both the angular distance from the lens
center {\it and} the azimuthal coordinate of the source as
measured with respect to the
symmetry axes of the lens.  At a given angular distance, $\theta$, from
the lens center, the magnitude of the shear is greatest for sources
located nearest to the major axis of the lens and least for sources located
nearest to the minor axis of the lens.  Hence, in a given radial annulus
which is centered on the lens, the
mean shear experienced by sources whose azimuthal coordinate, $\varphi$,
places them within $\pm N^\circ$ of the major axis of the lens  will be
greater than that for sources whose azimuthal coordinate, $\varphi$,
 places them within
$\pm N^\circ$ of the minor axis.  As a shorthand notation, we 
refer to the magnitude of the mean shear experienced by sources whose azimuthal
coordinates place them within $\pm N^\circ$ of the minor axis of the
lens as $\left<\gamma^- \right>$.  Similarly, we refer to the magnitude
of the mean shear experienced by sources whose azimuthal coordinates place
them within $\pm N^\circ$ of the minor axis of the lens as 
$\left< \gamma^+ \right>$.  For clarity,
we illustrate our use of this notation 
in Fig.\ \ref{diagram}.

For each galaxy in our simulations
we define an ``intrinsic'' shape parameter,
$\vec{\gamma}_i$, which describes
the image of the galaxy prior to being lensed:
\begin{equation}
\vec{\gamma}_i \equiv \frac{a-b}{a+b} e^{2i\phi},
\label{gammai}
\end{equation}
where $a$ and $b$ are the intrinsic major and minor axes of the light
distribution and $\phi$ is its intrinsic position angle.   Since we are again
dealing only with the weak lensing regime, the final shape of a galaxy,
$\vec{\gamma}_f$, after having
been lensed by $M$ foreground galaxies is given by:
\begin{equation}
\vec{\gamma}_f = \vec{\gamma}_i + \sum_{j=1}^M \vec{\gamma}_j,
\label{gammaf}
\end{equation}
where $\vec{\gamma}_j$ is the weak shear due to lens $j$.

For each galaxy, $i$, which has a
redshift of $z_i$, all 
other galaxies, $j$, with redshifts $z_j < z_i$ can formally 
be considered to be
lenses of galaxy $i$.   However,
since the galaxy--galaxy lensing signal drops off steeply with
projected lens--source separation on the sky (i.e., 
$|\vec{\gamma}(\theta)| \propto \theta^{-1}$ on scales, $\theta$, which are less
than or of order the angle subtended by the lens, and 
$|\vec{\gamma}(\theta)|$ falls off even more
steeply on scales larger than that subtended by the lens), 
we do not actually compute the weak shear due to all of the
foreground galaxies, $j$, which are within
the full $40' \times 40'$ region of the simulation.  Instead, to reduce
the run time of the simulations, we simply compute the
net shear on each (source) galaxy $i$ due to all (lens) galaxies $j$  which are
located within $150''$ of galaxy $i$.  Increasing
the maximum lens--source separation to $300''$ has a negligible effect
($< 0.01 | \vec{\gamma}_f|$) 
on the net shear computed for each of the source galaxies.

\section {Results}

Using the final image shapes of the galaxies, $\vec{\gamma}_f$, we
compute the angular dependence of the mean value of the shear 
experienced by sources located closest to the minor axes of 
lens galaxies, $\left< \gamma^- \right>$, as well as the mean value
of the shear experienced by sources located closest to the major
axes of lens galaxies, $\left< \gamma^+ \right>$ (e.g.,
Fig.\ \ref{diagram}).  Clearly, in the case of isolated lenses for which
the position angle of the {\it mass} distribution of the lens (i.e., the
position angle of the halo as projected on the sky) is known precisely,
we expect $\gminor < \gmajor$ on all angular scales, $\theta$, which are
less than or of order the angular scale subtended by the halos.  Of course,
on angular scales which are substantially
larger than that subtended by the halos, the
ratio $\gminor / \gmajor$ will approach unity since, for all intents
and purposes, the truncated elliptical lens begins to act similarly to
a point mass lens.

Here we restrict our analysis to galaxies 
for which $19 < I < 23$, since a limiting magnitude of
order $I=23$ can be achieved in a reasonable amount of observing
time with large ground--based telescopes, and because this limiting 
magnitude is comparable to the deeper data sets which have been used to 
detect galaxy--galaxy lensing.  We do, however, include
the effects of systematic lensing due to galaxies with $23 \le I \le 25$
on the images of galaxies with $I < 23$
since the redshifts of some galaxies
with $I \ge 23$ will, in fact, be smaller than the redshifts of some
galaxies with $I < 23$ (e.g., equation (\ref{zdist}) above).
In addition, because of the finite area of the fields (1600 square arcminutes
each), we restrict our computation of $\gminor$ and $\gmajor$
to only those galaxies which are 
located within
the central $30' \times 30'$ region of each simulation.  This is done
in order to avoid errors in the calculation of the mean shear which would occur
simply due to boundary effects caused by the fact that 
source galaxies located
close to the edges of the simulations will not have been
lensed by objects which would, in a larger simulation, exist just outside
our $40' \times 40'$ frame.

In all cases below, we compute the mean shear only on scales
greater than $\theta = 5''$.  This is identical to the inner radius
imposed by BBS in their analysis and it insures that in a realistic
data set the ``source'' galaxies in the analysis will be located more
than 7 scale lengths from the ``lens'' galaxies (see, e.g., BBS).  This,
in turn, insures that systematic errors due to overlapping isophotes
of lens and source galaxies should not occur in a realistic observational
data set.

\subsection{No Observational Noise}

We first compute the angular dependence of $\gminor$ and $\gmajor$ in
the limit of truly ``ideal'' data.  That is, from the simulations
we have precise knowledge of
the orientation of the major axes of all of the halos, the
redshifts of the galaxies, and the final shape parameters, $\vec{\gamma}_f$,
of the images of the galaxies after they have been lensed by all foreground
galaxies.
The only source of ``noise'' in
the data, therefore,
is simply that due to the intrinsic ellipticity distribution of
the galaxy shapes.

\begin{figure}
\plotone{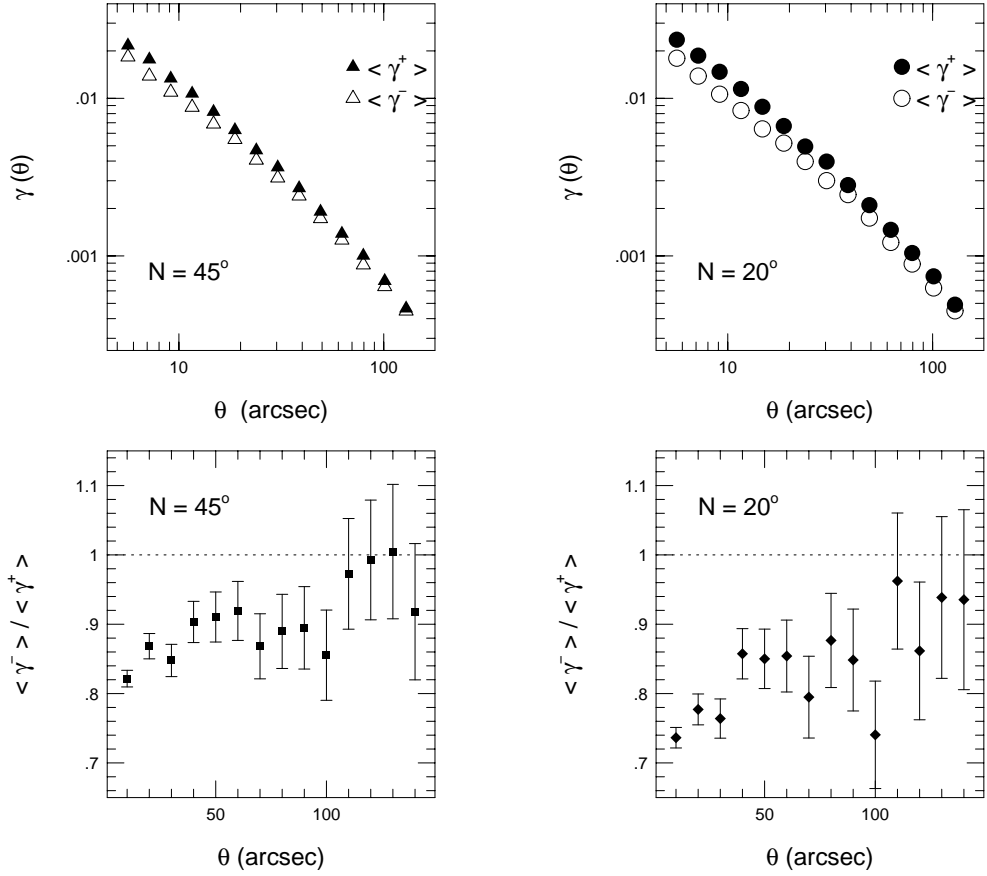}
\vskip -3.0in
\caption{
The dependence on differential
lens--source separation, $\theta$, of $\gmajor$, $\gminor$ (top
panels) and $\gminor / \gmajor$ (bottom panels).  Left panels are
for the case that all sources have azimuthal coordinates,
$\varphi$, which are within $\pm 45^\circ$ of the symmetry axes of
the projected halo mass.  Right panels are for the case that
all sources have azimuthal coordinates, $\varphi$, which are within
$\pm 20^\circ$ of the symmetry axes of the projected halo mass.
Here, all galaxies with $19 < I < 23$ are used in the
calculation.
\label{diff1}
}
\end{figure}

For all foreground galaxies, $j$, located at an angular distance $\theta$ from
all background galaxies, $i$ (i.e., $z_j < z_i$), we compute the net
tangential shear of the background galaxies using the individual values of 
$\vec{\gamma}_f$ which were
obtained in the simulations.  For all background galaxies
located within $\pm N^\circ$ of the symmetry axes of the halos, then,
we compute the angular dependence of $\gminor$, $\gmajor$, and the
ratio $\gminor / \gmajor$ as a function of differential
lens--source separation on the sky.  The results are shown 
in Fig.\ \ref{diff1}, where the
left panels show results for $N = 45^\circ$, and
the right panels show results for $N = 20^\circ$. 

As expected, the variation of $\gminor$ and $\gmajor$ with 
differential lens--source separation
is well--fitted by a broken power law.
That is, on scales, $\theta$, which are less than or of order the angle
subtended by the lenses, the mean shear should decrease roughly
as $\theta^{-1}$ since we have adopted an isothermal density profile.
On scales larger than the truncation radius, a steepening in the
variation of the mean shear with angular separation ought to occur and
this, indeed, is the case.  The changes in slope which are
shown in the top panels
of Fig.\ \ref{diff1} occur at 
$\theta \simeq 33''$ which, unsurprisingly, corresponds to the mean
projected value of the truncation radii of the halos,
$\left< x_t \right> \simeq 33''$.  For sources located within 
$\pm 45^\circ$ of the symmetry axes of lenses, the change in slope
is of order $-0.5$ (i.e., steeper than the slope for an isothermal
lens, but not quite as steep as that for a point mass lens).
Again as expected, the variation of the ratio $\gminor / \gmajor$ 
as a function of differential lens--source separation
is less than unity on small angular scales and approaches unity on
scales considerably larger than the mean projected value of the halo
truncation radii.  

\begin{figure}
\plotone{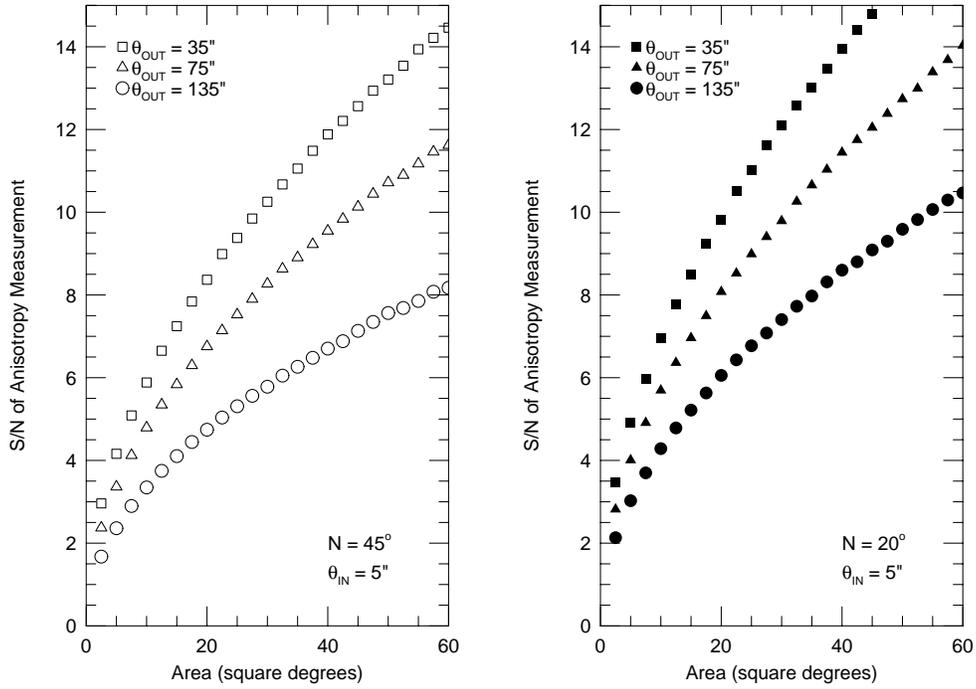}
\vskip -4.0in
\caption{The signal--to--noise ratio which can be achieved for
a measurement of the anisotropy parameter, ${\cal A}$, for the
case of no observational noise.  The left panel shows the results for
$N=45^\circ$ (measured relative to the intrinsic position angles of
the lens galaxies) and the right panel shows the results for 
$N=20^\circ$.  Results are shown for three fiducial analysis annuli:
$5'' \le \theta \le \theta_{\rm out}$, $\theta_{\rm out} = 35''$
(squares), $75''$ (triangles), and $135''$ (circles).
\label{sn1}
}
\end{figure}

For the purposes of planning possible future
observational investigations, we are interested in estimating
the size of a deep imaging
survey which would be required to detect the effects of flattened
halos on the galaxy--galaxy lensing signal.
To do this we define an ``anisotropy parameter'',
\begin{equation}
{\cal A} \equiv 1 - \frac{\gminor}{\gmajor}, 
\end{equation}
and, again for the case
of truly ideal data, we compute the signal--to--noise in a measurement
of ${\cal A}$ as a function of the area of the survey.  Shown in 
Fig.\ \ref{sn1} are the results for lens--source separations of
$5'' \le \theta \le \theta_{\rm out}$, where $\theta_{\rm out} = 
35'', 75''$, and $135''$ for the cases of $N = 45^\circ$ (left panels)
and $N = 20^\circ$ (right panels).  For the case of $N = 20^\circ$,
the values of the anisotropy parameter are ${\cal A} = 0.24, 0.20,
0.18$ for $\theta_{\rm out} = 35'', 75'', 135''$ and a $4\sigma$ 
detection of ${\cal A}$ should be obtained with surveys of area
3.5, 5.0, and 9.0 square degrees, respectively.
For the case of $N = 45^\circ$, the values of the anisotropy parameter
are ${\cal A} = 0.15, 0.13, 0.11$ for $\theta_{\rm out} = 35'', 75'', 135''$
and a $4\sigma$ detection of ${\cal A}$ should be obtained with surveys of
area 5.0, 7.0, and 14.5 square degrees, respectively.

\subsection{Lensing--Induced Image Correlations}

In the above subsection, we computed $\gminor$ and $\gmajor$ 
relative to the 
actual symmetry axes of the halos of the lens galaxies themselves.  That is,
$N$ was specifically measured relative to the major and minor axes of the
{\it halos} as seen in projection. 
Of course, in a real data set one must measure $N$ relative to the 
major and minor axes of the {\it light} distribution of the lens galaxies
and, to some extent, the position angle of a lens galaxy's equivalent image
ellipse will differ from its intrinsic position angle due to the fact
that lens galaxies with redshifts $z_i$ will have been weakly--lensed by
foreground galaxies with redshifts $z_j < z_i$.  For a typical galaxy in
our simulation, the misalignment between the intrinsic position angle of
the galaxy and its final post--lensing position angle is of order 
$5^\circ$ or less.  This, clearly, is a source of intrinsic noise in the
determination of $N$ which one might think would be very small indeed
(particularly for values of $N >> 5^\circ$).  However, as we show
below, this slight deflection of the position angles of the lens galaxies
away from their intrinsic position angles (i.e., those which were aligned
with the halo major axes {\it a priori}) induces a substantial systematic error
in the values of $\gminor$ and $\gmajor$ when these quantities are
computed relative to the post--lensing position angles of the images
of the lens galaxies.

\begin{figure}
\plotone{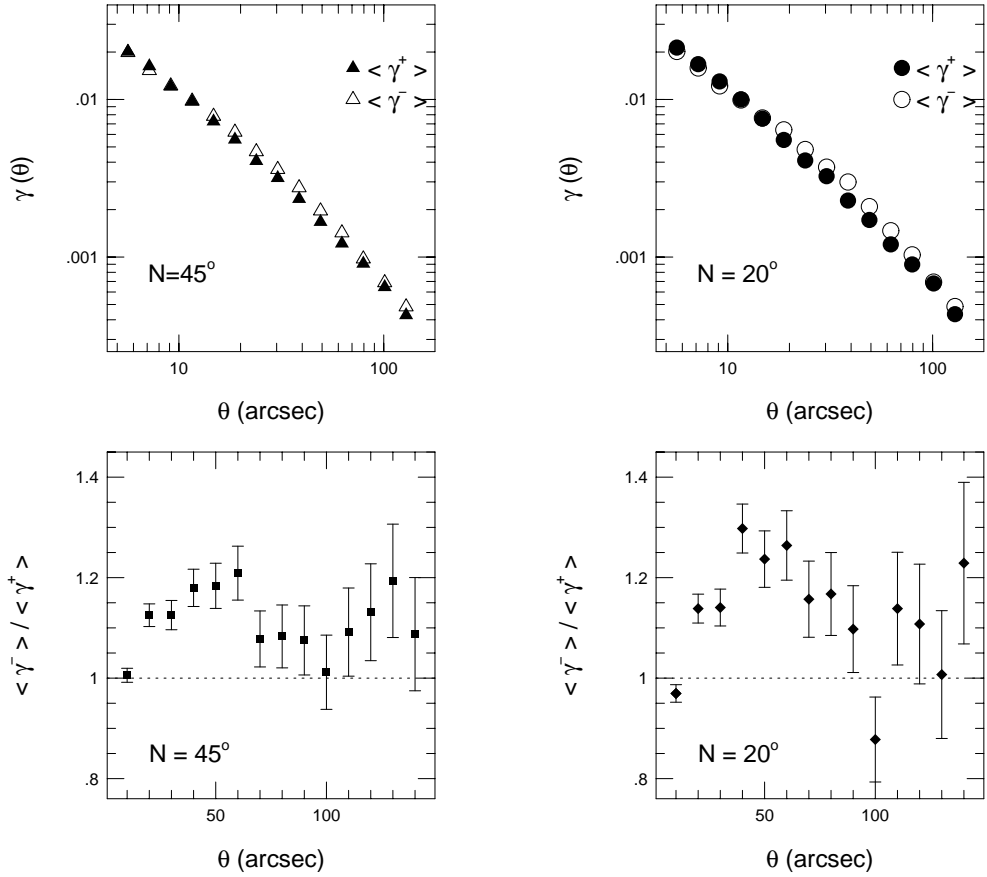}
\vskip -3.0in
\caption{
Same as for Fig.\ \ref{diff1}, but here $N$ is measured relative to
the final, post--lensing position angles of the images of the
lens galaxies.
\label{diff2}
}
\end{figure}

Shown in Fig.\ \ref{diff2}
is the angular dependence of $\gminor$,
$\gmajor$, and $\gminor / \gmajor$ for $N = 45^\circ$ (left panels)
and $N = 20^\circ$ (right panels), where $N$ has been measured relative
to the final, observed position angles of the lens galaxies.  That is,
the sole difference between Fig.\ \ref{diff2} and
Fig.\ \ref{diff1} is that here $N$ is measured relative
the final, post-lensing position angles of the images of the lens galaxies,
whereas previously we measured $N$ relative to the intrinsic position
angles of the lenses.  In Fig.\ \ref{diff2} 
we see quite a striking trend:
on scales larger than about $10''$ the values of $\gminor$ actually exceed
those of $\gmajor$ and, hence, the ratio $\gminor / \gmajor$ is greater
than unity.  

In the limit of single deflections for each source galaxy, and for the
case of isolated lens galaxies which are not themselves lensed by any
foreground object, we always expect $\gminor / \gmajor$ to be less than
unity on small scales.  However, due to the fact that our galaxies
are broadly distributed in redshift and we have taken full account of all
weak lensing events for all of the galaxies, we clearly do not have
this idealized limiting case.  Rather, consider the case of 3 galaxies with
redshifts $z_1 < z_2 < z_3$.  If the galaxy at $z_3$ is weakly--lensed by both
of the galaxies at $z_2$ and $z_1$, 
and the galaxy at $z_2$ is also weakly--lensed
by the galaxy at $z_1$, then correlated alignments of the post--lensing images
of the galaxies at $z_2$ and $z_3$ will occur.  This, in turn, will
cause $\gminor / \gmajor$ as measured relative to the final image of the
lens at $z_2$ to be
greater than unity (effectively by decreasing the net ellipticity of sources
at $z_3$ which are closest to the projected major axis of the halo of 
the lens at $z_2$ and, simultaneously, increasing the net ellipticity of
sources at $z_3$ which are closest to the minor axis of the halo).

We quantify this effect in our simulations using a
correlation function of the image shapes,
\begin{equation}
C_{\gamma \gamma} (\theta) \equiv \left< \vec{\gamma}_{1} \cdot
\vec{\gamma}_{2}^\ast \right>
\label{eq_cpp}
\end{equation}
where the mean value is computed for all foreground--background pairs of
galaxies separated by angles $\theta \pm \delta\theta/2$ 
on the sky (see, e.g.,
Blandford et al.\ 1991). 
Here $\vec{\gamma}_{1}$ is the image shape of a galaxy with 
redshift $z_1$ and $\vec{\gamma}_{2}^\ast$ is the complex conjugate
of the image shape of a galaxy
with redshift $z_2 > z_1$ (see, e.g., equations \ref{gammai} and 
\ref{gammaf} above).  
This function measures the degree to which the
images of galaxy 1 and galaxy 2 are aligned with one another and in the 
limit of no systematic alignments between images 1 and 2, $C_{\gamma \gamma}
(\theta)$ is identically zero on all scales.

\begin{figure}
\vskip -0.75in
\plotone{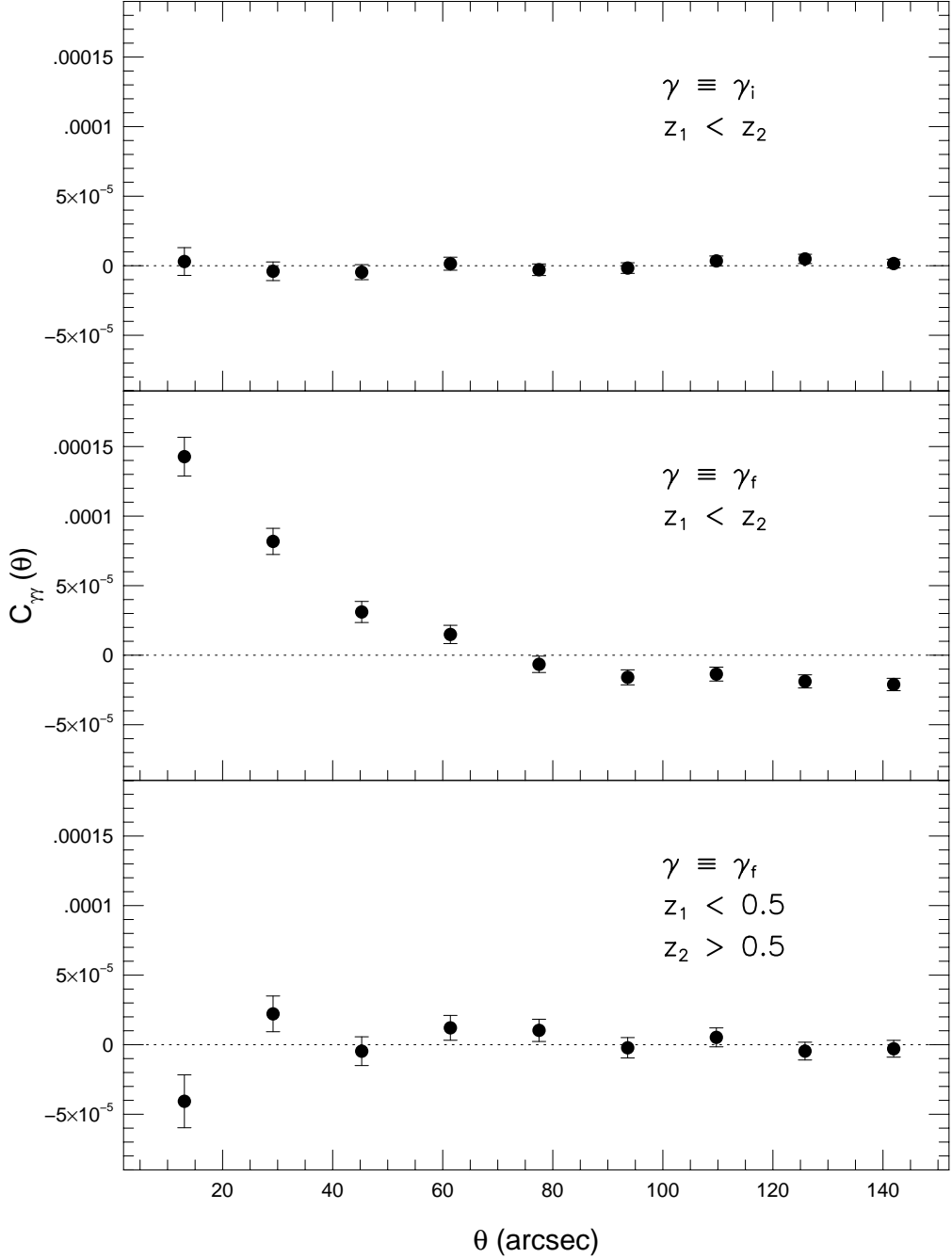}
\vskip -0.75in
\caption{The correlation function of image shapes for
Monte Carlo galaxies with $19 < I < 23$; i.e., equation (\ref{eq_cpp}) .
Top panel: $C_{\gamma\gamma}$ computed using the intrinsic shape parameters,
$\vec{\gamma}_i$,
which were drawn at random at the start of each simulation.  Middle
panel: $C_{\gamma\gamma}$ computed using the final shape parameters,
$\vec{\gamma}_f$, after all galaxies had been lensed by all foreground
galaxies. Bottom panel: $C_{\gamma\gamma}$ computed using the final shape
parameters for the specific case in which  $z_1 < 0.5$ and $z_2 > 0.5$.
\label{cpp}
}
\end{figure}

Shown in the top panel of Fig.\ \ref{cpp} is the function 
$C_{\gamma \gamma}(\theta)$,
computed using the intrinsic shape parameters, $\vec{\gamma}_i$, of 
all galaxies
with magnitudes in the range $19 < I < 23$.  As expected, 
$C_{\gamma \gamma}(\theta)$ 
is consistent with zero on all scales since all galaxies
are assigned intrinsic shape parameters which are drawn at random
at the start of each simulation.  Shown
in the middle panel of Fig.\ \ref{cpp} is $C_{\gamma \gamma}(\theta)$ 
computed using
the final shape parameters for all galaxies with $19 < I < 23$ (i.e., the
objects which we have used to measure $\gminor$ and $\gmajor$ in 
Figs.\ \ref{diff1} and \ref{diff2}).  Here there is
clearly a net correlation of foreground--background image shapes
on scales $\theta \lo 60''$ and a slight anti--correlation on larger scales.  It
is this small--scale correlation of the final image shapes between 
foreground and background galaxies which is responsible for the ratio
$\gminor / \gmajor$ being greater than unity in Fig.\ \ref{diff2}.
That is, the effect of multiple weak lensing events on {\it pairs}
of lenses and sources is to cause a slight tendency for the images of
lenses and sources to be preferentially aligned with each other on angular
scales $\lo 60''$ in our simulations.

This effect will, of course, be strongest for foreground--background pairs
in which the redshifts of the foreground and background objects are fairly
similar (i.e., each of the two objects is likely to be multiply--lensed by
an identical number of foreground objects).  Therefore, it should be
possible to reduce the effect of correlated lens--source shape parameters by
restricting our analysis to lens redshifts, $z_d$, which are relatively
smaller than the source redshifts, $z_s$.  We find that this does, indeed,
occur when we restrict our analysis to the following cases:
(i) $z_d < 0.1$, $z_s > 0.1$;
(ii) $z_d < 0.2$, $z_s > 0.2$;
(iii) $z_d < 0.5$, $z_s > 0.5$. 
Redshift cuts using $z_d < 0.1$ and $z_d < 0.2$ result in values of
$\gminor / \gmajor$ which are less than unity, but which require 
excessively wide
fields in order to detect our anisotropy parameter, ${\cal A}$, at a 
significant level.  A redshift cut at $z_d < 0.5$, however, largely 
eliminates any correlation between the values of $\vec{\gamma}_f$ for all 
foreground--background pairs of galaxies (see the bottom panel of Fig.\ 
\ref{cpp}) and still allows a significant detection of ${\cal A}$ with
only moderate--sized data sets.

\subsection{Effects of Observational Noise}

In the above sections we have demonstrated that in the limit of imaging
data which contains no observational noise whatsoever, it is possible
in principle to detect the effects of flattened halos on the
galaxy--galaxy lensing signal.  That is,
the only noise for which we have accounted
so far is the ``intrinsic'' noise associated with the fact that the
galaxy images have a broad distribution of intrinsic ellipticities and 
that multiple weak lensing events can cause the images of foreground--background
pairs of galaxies to be preferentially aligned with one another.

Here we include the effects of realistic observational noise
in our analysis in order to determine whether or not a detection of
galaxy--galaxy lensing by flattened halos would be feasible with
ground--based imaging data of modest quality.  It is unlikely
that spectroscopic redshifts would be available for all of the galaxies
in an observational data set which has a depth comparable to our simulations;
however, photometric redshifts could certainly be obtained in a
multi--color data set
and these could be used for the purposes of lens--source separation.
Photometric redshifts are, of course, not exact and this will introduce
some level of noise into the analysis.
In addition, seeing, sky noise, and pixellation effects introduce noise
into the observed, final image shapes (i.e., the observed
values of $\vec{\gamma}_f$)
and they, therefore, degrade the galaxy--galaxy lensing signal itself.

We model the effects of noise which would be
introduced by the use of photometric
redshifts (as opposed to spectroscopic redshifts) by adding an error term 
to each galaxy's actual redshift:
\begin{equation}
z_{\rm obs} = z_{\rm in} + \delta z,
\end{equation}
where $z_{\rm in}$ is the intrinsic redshift assigned to the galaxy
at the start of the simulation and 
$\delta z$ is an error term drawn from a Gaussian distribution with zero mean
and a standard deviation of 0.1.  This is in reasonable agreement with the
accuracy of photometric redshifts found by Hogg et al.\ (1998) who 
conducted a blind test of the predictions of several photometric redshift
algorithms against the actual spectroscopic redshifts of galaxies in the
HDF--N and found that for 68\% of the galaxies the error in
the photometric redshift was $\lo 0.1$.

We model the effects of noise introduced in the imaging process 
by adding appropriate errors to the final, observed position angles of
the galaxies.  This is identical to the approach followed by BBS in
their simulations of galaxy--galaxy lensing and is motivated
to a large extent by simplicity.  That is, although there will certainly
be a finite error associated with the measurement of the ellipticity of
a galaxy's image, we choose to incorporate all of the observational error
in a measurement of $\vec{\gamma}_f$ into an error in the final, observed
position angle since it is largely the degree of tangential alignment
of the source galaxies relative to the lens galaxies which gives rise to
the significance of a detection of galaxy--galaxy lensing. 

The magnitudes of the galaxies used by BBS in their observational
investigation of galaxy--galaxy lensing, $20 \le r \le 24$,
were very similar to those
adopted for our simulations and we use the results of 
matched galaxy catalogs from substacks of the BBS data set to assign
errors in the position angles of the galaxies in our simulations.
From substacks of the data, BBS noted that the typical error in the
position angle of a faint galaxy is dependent upon not only the magnitude
of the galaxy but also its ellipticity, with the error being most 
significant for the roundest galaxies (see, e.g., Table 2 of BBS).  We
extend the original analysis of BBS to galaxies which are brighter than those
which are tabulated in their paper 
and in Table 1 we show the typical errors
in the position angles of galaxies with $19 < I < 23$ as a function of
both $I$--magnitude and ellipticity, $\epsilon$.  We also note that the
median seeing in the BBS data was $0.9''$ FWHM and no attempt was made 
by BBS to deconvolve the images, 
so the data should be fairly representative of the imaging
quality which could be achieved under modest observing conditions.

For each galaxy in the simulation, then, we assign an error to the
final, observed (i.e., post--lensing) position angle based upon the
galaxy's apparent magnitude and its ellipticity after having been lensed
by all foreground objects.  The error is assigned from a Gaussian distribution
with zero mean and the appropriate standard deviation from Table 1.  Thus,
we now have a set of objects with redshifts ($z_{\rm obs}$) and
final shape parameters 
($\vec{\gamma}_f^{\rm obs}$) which have been ``corrupted''
by the inclusion of moderate errors.

\begin{figure}
\plotone{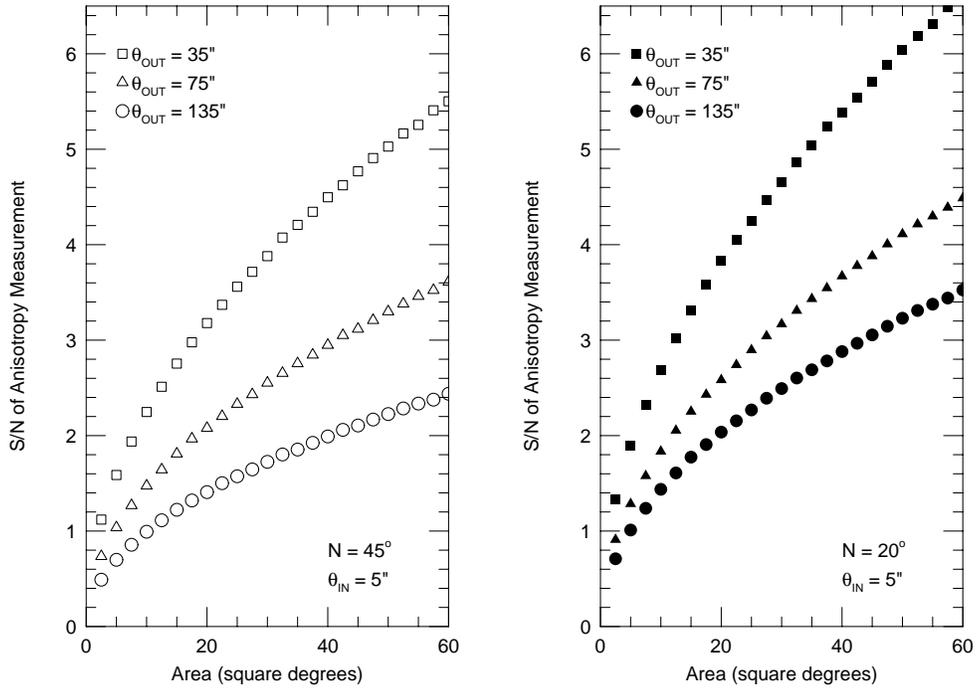}
\vskip -4.0in
\caption{Same as Fig.\ \ref{sn1},
but here $N$ is measured relative
to the final, post--lensing position angles of the images of the
foreground galaxies and noise has been added to both the
redshifts of the galaxies and the values of the observed position
angles (see text and Table 1). Here 
we restrict the analysis to foreground
galaxies with $z_d^{\rm obs} < 0.5$ and background galaxies 
with $z_s^{\rm obs} > 0.5$ in
order to break the correlation in the values of $\vec{\gamma}_f^{obs}$ for 
foreground--background pairs. 
\label{sn3}
}
\end{figure}

Using these corrupted redshifts and final shape parameters, then, we
re--compute the signal--to--noise which can be achieved in a measurement
of our anisotropy parameter, ${\cal A}$, as a function of the area of
the survey.  Again, to reduce the effects of correlated shape parameters
of foreground and background galaxies, we use only lens galaxies with
$z_d^{\rm obs} < 0.5$ and source galaxies with $z_s^{\rm obs} > 0.5$ 
in the analysis.  The results are shown in Fig.\ \ref{sn3} where, as
in Fig.~\ref{sn1}, the left panel corresponds to the case
in which  $N = 45^\circ$ and the right panel corresponds to the case in
which $N = 20^\circ$.  Not unexpectedly, we find that for a given survey area
the signal--to--noise in a measurement of ${\cal A}$ decreases significantly
when moderate amounts of observational noise are included in the 
calculation.  Nevertheless, when the signal is
averaged over relatively small angular scales, $5'' \le \theta \le 35''$,
a 4$\sigma$ detection of ${\cal A}$ can be achieved with a data set
of order 22 square degrees in the case of $N = 20^\circ$ and with a data set
of order 32 square degrees in the case of $N = 45^\circ$. 

\subsection{Weighted Means}

In the previous sections we have used only simple, unweighted means
in our calculations of $\gminor$ and $\gmajor$.  However, as our two
illustrative cases of $N = 20^\circ$ and $N = 45^\circ$ demonstrate, the
``signal'' (i.e., the level of anisotropy) is greatest for sources
located closest to the symmetry axes of lenses and will, of course, be
nearly identical for all sources which have azimuthal coordinates, $\varphi$,
which are close to $45^\circ$.  In principle, therefore, one should consider
computing $\gminor$ and $\gmajor$ using all sources within $N = 45^\circ$
via a weighted mean which would weight sources with $N << 45^\circ$ strongly,
while largely discarding sources with $N \simeq 45^\circ$ from the calculation.

\begin{figure}
\vskip -0.75in
\plotone{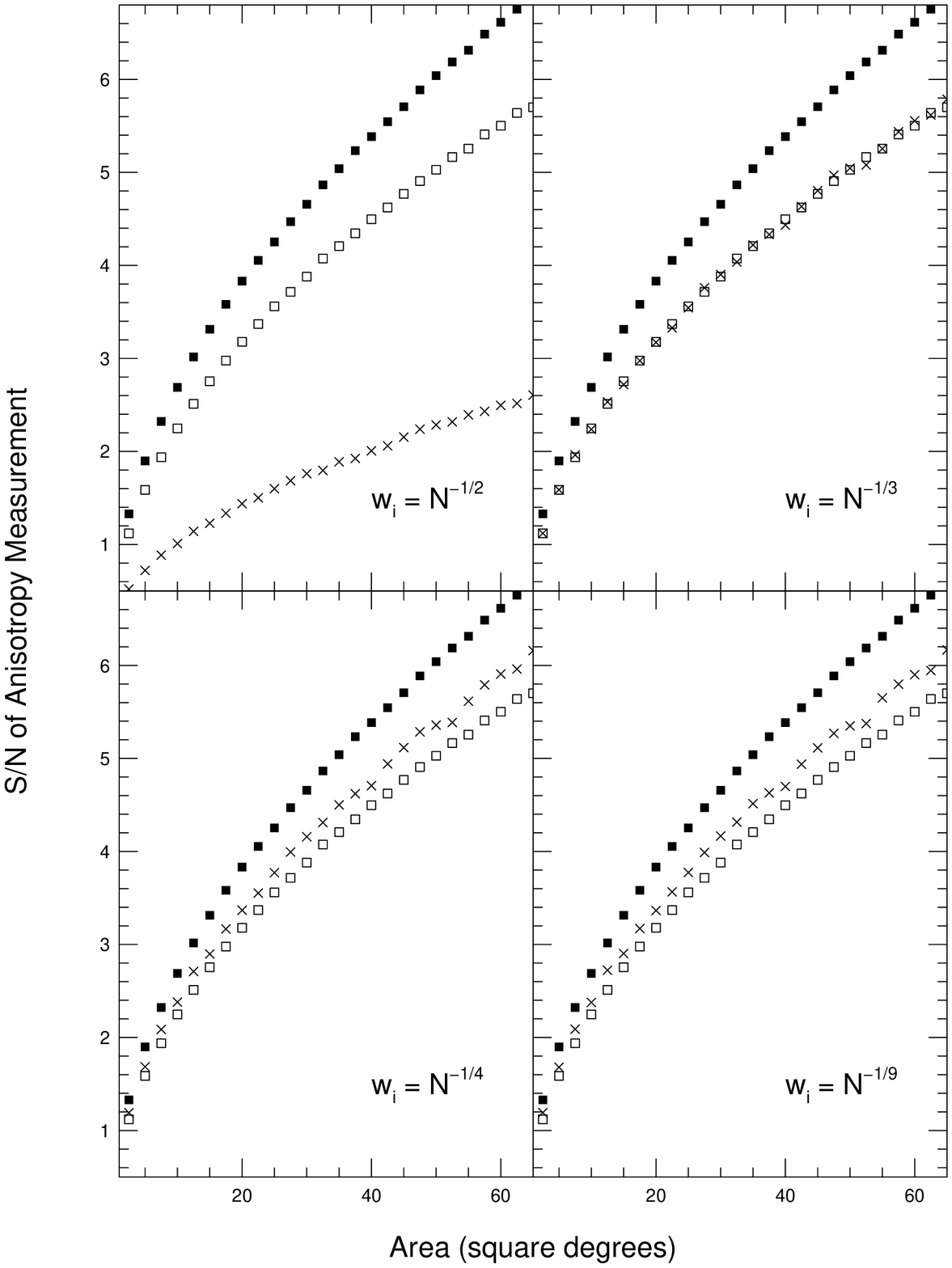}
\vskip -0.75in
\caption{Signal--to--noise as a function of survey area for noisy
data.  Here the signal is averaged over angular scales
of $5'' \le \theta \le 35''$ and the azimuthal coordinates of the sources
are measured relative to the final, post--lensing position angles of the
images of the lenses to which noise has been added. 
As in Fig.\ \ref{sn3} we have restricted the 
analysis to foreground
galaxies with $z_d^{\rm obs} < 0.5$ and background galaxies
with $z_s^{\rm obs} > 0.5$ in
order to break the correlation in the values of $\vec{\gamma}_f$ for
foreground--background pairs. Squares show S/N for simple, unweighted
means (solid symbols for $N = 20^\circ$, open symbols for $N = 45^\circ$).
Crosses show S/N for weighted means in which the weights are given
by $N^{-1/2}$ (top left panel), $N^{-1/3}$ (top right panel),
$N^{-1/4}$ (bottom left panel), and $N^{-1/9}$ (bottom right panel), where
$N \le 45^\circ$.
\label{sn4}
}
\end{figure}

In this section we compute $\gminor$ and $\gmajor$ using simple weighted
means where the weight applied to the image of galaxy $i$ is of the form:
\begin{equation}
w_i = N^{- 1/n},
\end{equation}
where $n=2, 3, 4, ... 9$.  For brevity, we show only results for the
case of the noisy data used in \S4.3 
(i.e., noise added to both the image shapes
and the redshifts), and a fiducial analysis annulus of 
$5'' \le \theta \le 35''$.  Results for signal-to-noise as a function
of survey area as obtained from the weighted means are shown in
Fig.\ \ref{sn4}, where they are directly compared to the results of the 
previous unweighted means (i.e., Fig.\ \ref{sn3}).

From Fig.\ \ref{sn4}, it is clear that when all sources within 
$N = 45^\circ$ of the lens symmetry axes are used, this weighting scheme
yields an improvement in the S/N over the unweighted mean for the case
of $n > 4$.  However, increasing $n$ beyond a value of 4 does not increase
the S/N substantially for a given survey area and, in fact, the simple
unweighted mean which uses all sources within $N = 20^\circ$ of the lens
symmetry axes yields the largest signal to noise values.

\section{Discussion}

We have used the well--studied HDF--N to show that the probability of
multiple weak deflections due to galaxy--galaxy lensing
is significant in a deep data set ($I_{\rm lim} \sim 23$).  Using a 
fiducial model in which the halos of $L^\ast$ galaxies have characteristic
velocity dispersions of 156~km/s and outer radii of 100$h^{-1}$~kpc,
we have shown that neglecting multiple weak deflections when predicting
the galaxy--galaxy lensing shear field results in a substantial
underestimate of
the mean shear on scales $> 1''$.  This, in turn,
leads to an overestimate of the mass of the halo of an $L^\ast$ galaxy
by a factor of order 2 interior to a radius of 100$h^{-1}$~kpc.

We have also used Monte Carlo simulations of
galaxy--galaxy lensing by flattened dark matter halos
to compute the signal--to--noise which might be
achieved for a measurement of ${\cal A} \equiv 1 - \gminor / \gmajor$.
For modest degree of observational noise (including errors in photometric
redshifts and in the determination of the final image shapes) in a data set
for which the galaxy magnitudes are in the range $19 < I < 23$, we find that
a simple unweighted mean in the computations of $\gminor$  and $\gmajor$
should yield a significant detection of the weak lensing signal due to flattened
halos provided the data set is sufficiently wide.  Multiple weak deflections
(i.e., weak lensing of background lens--source pairs by foreground galaxies)
do, however, tend to cause alignments between the final
images of lenses and sources,
which complicates the detection of the expected anisotropy in the 
galaxy--galaxy lensing signal.  A simple redshift cut of $z_d < 0.5$,
$z_s > 0.5$ is sufficient to break this correlation, however.  Imposing
such a redshift cut, and restricting our unweighted mean to sources within
$N= 20^\circ$ of the lens symmetry axes, 
we find that a 4$\sigma$ detection of the
anisotropy parameter can be obtained with a data set of order 22 square 
degrees.  If all sources within $N= 45^\circ$ of the lens symmetry axes
are used, the required survey area increases to of order 32 square degrees
for a 4$\sigma$ detection. 

Two previous investigations, Natarajan \& Refregier (2000) and Brainerd
\& Wright (2000), have also made predictions for the sizes of data sets
which would be necessary in order to detect the anisotropic
signature of systematic
weak lensing by flattened dark matter halos.  Neither of these investigations
included the effects of multiple weak deflections on the lensing signal and both
considered only the very simple case in which all galaxy halos are
modeled as infinite singular isothermal ellipsoids with identical
projected ellipticities, $\epsilon$.  

Brainerd \& Wright (2000) used the observed signal--to--noise in the
galaxy--galaxy lensing signal obtained by BBS to estimate the size of
a data set with imaging characteristics similar to the BBS data which
would be required in order to detect galaxy--galaxy lensing by flattened halos.
They computed the anisotropy parameter, ${\cal A}$, for the
case of $N = 45^\circ$ and used unweighted means in the computations of
$\gminor$ and $\gmajor$.  For halo mass ellipticities of $\epsilon = 0.3$, and
using all sources within $N = 45^\circ$ of the lens symmetry axes,
Brainerd \& Wright (2000) concluded that a mere 1.25 square degrees of imaging
data similar to that of BBS would be sufficient to yield a 4$\sigma$ detection
of weak lensing by flattened halos when averaged over lens--source 
separations of $5'' \le \theta \le 35''$.  This estimate is, however, a 
factor 
of order 25 too small compared to the results of the simulations discussed
here
(i.e., left panel of Fig.\ \ref{sn3}).

Natarajan \& Refregier (2000) used a complimentary approach to predict that,
in the absence of observational noise, the signal--to--noise which could
be achieved in a measurement of:
\begin{equation}
\epsilon_{\kappa} = \frac{a^2 - b^2}{a^2 + b^2}
\end{equation}
is given by:
\begin{equation}
\left( \frac{S}{N} \right)_{\epsilon_\kappa} \simeq
4.6 \left( \frac{\epsilon_\kappa}{0.3} \right)
\left( \frac{\alpha}{0.5''} \right)
\left( \frac{n_b}{1.5~{\rm arcmin}^{-2}} \right)^{1/2}
\left( \frac{n_f}{0.035~{\rm arcmin}^{-2}} \right)^{1/2}
\left( \frac{0.3}{\sigma_\epsilon} \right)
\left( \frac{A}{1000~{\rm deg}^2} \right)^{1/2}
\label{sn_priya}
\end{equation}
where $\alpha$ is the Einstein radius, $n_b$ is the number of background
galaxies, $n_f$ is the number of foreground galaxies, $\sigma_\epsilon^2$
is the variance in the intrinsic ellipticity distribution of galaxies
($\sim 0.3^2$), and $A$ is the area of the survey.
For a data set similar to that of BBS (i.e., similar to the type of data set
investigated in this paper and in Brainerd \& Wright 2000), we have that
$\alpha \sim 0.17''$, $n_b = 7~{\rm arcmin}^{-2}$, and
$n_f = 6~{\rm arcmin}^{-2}$ (see, e.g., BBS).  Adopting these values of
$\alpha$, $n_b$, and $n_f$ as being representative, then, equation 
(\ref{sn_priya})
predicts that for a value of $b/a = 0.7$ (i.e., $\epsilon = 0.3$ in our
notation and $\epsilon_\kappa = 0.34$ in the notation of Natarajan
and Refregier), a 4$\sigma$ detection of weak lensing by flattened halos
should be detectable in a survey with an area of order 6 square degrees.
Again, compared with the predictions of our simulations, this is a considerable
underestimate of the size of the data set which would be needed in order
to detect the anisotropic lensing signal at a significant level. 

The primary motivation for studies of galaxy--galaxy lensing has, of course,
been its promise to place strong statistical constraints on the physical
parameters of the halos of galaxies.  This motivation is well--justified
because of the fact that weak lensing provides a probe of 
the gravitational potentials of the halos of the
lens galaxies up to very large physical radii (on the order of
$100h^{-1}$~kpc or so), where dynamical and hydrodynamical tracers of
the potential are very unlikely to be found for the vast majority of 
the galaxies in the universe.  To date, the published studies of
galaxy--galaxy lensing by field galaxies have attempted
to constrain the depths of the potential wells of the halos of 
the lens galaxies and their typical outer scale radii, but they are only
just beginning to
address the shapes of the halos.  In a very preliminary, but nevertheless
tantalizing, study, Hoekstra, Yee \& Gladders (2002) 
have used a maximum likelihood
technique to constrain the shapes of the halos of early--type galaxies
in the Red--Sequence Cluster Survey
and they find that they can rule out spherical halos at the 99\% confidence 
level. Although not yet a definitive result, this is certainly encouraging
for future studies which aim to constrain halo shapes via weak lensing.

Here we have explored galaxy--galaxy lensing by field galaxies with
flattened halos and we have concluded that the resulting weak lensing
signal should be detectable in deep, wide, multi-color data sets with
modest imaging quality.  That is, provided the halos have a median
ellipticity of order 0.3, we find that an anisotropy in the weak lensing
signal (i.e., $\gminor$ vs.\ $\gmajor$) is detectable via an appropriate
strategy in the data analysis.  However, in closing we caution that this is not
identical to a direct measurement of the mean flattening of the 
halo population; it is, rather, an indicator that the halos are not
spherically--symmetric on average.  If all galaxy halos were infinite
singular isothermal ellipsoids with identical ellipticities in projection
on the sky, and all galaxy
images were weakly--lensed by one and only one foreground galaxy, then it
would be very straightforward to turn a single measurement of 
$\gminor / \gmajor$,
or equivalently ${\cal A}$, into a direct measurement of $\epsilon$ (see, e.g.,
Brainerd 2002 and Brainerd \& Wright 2000).  It is, of course, not realistic
to suppose that all halos have identical ellipticities in projection and
this, combined with the need to remove lensing--induced
correlations between the
final shapes of lenses and sources will
certainly necessitate the use of detailed Monte Carlo simulations to translate
any future measurements of 
anisotropies in the galaxy--galaxy lensing signal into 
constraints on the mean flattening of the halos of field galaxies.

\section*{Acknowledgments}

Support under NSF contracts AST-9616968 (TGB and COW), AST-0098572 (TGB), 
an NSF Graduate  Minority
Fellowship (COW), and a generous allocation of resources
at Boston University's Scientific Computing and Visualization 
Center are gratefully acknowledged.

\clearpage

\centerline{    }
\vskip 2.0in
\centerline{{\bf Table 1: Errors Assigned to Post--Lensing Position Angles}}

\vspace{0.1in}

\centerline{
\begin{tabular}{ccccc} 
$\epsilon$ & $19<I<20$ & $20<I<21$ & $21<I<22$ & $22<I<23$ \\ \hline
0.0--0.1 & 30$^{\circ}$ & 30$^{\circ}$ & 30$^{\circ}$ & 35$^{\circ}$ \\ 
0.1--0.2 & 13$^{\circ}$ & 16$^{\circ}$ & 20$^{\circ}$ & 25$^{\circ}$ \\ 
0.2--0.3 & 8$^{\circ}$ & 11$^{\circ}$ & 15$^{\circ}$ & 20$^{\circ}$ \\ 
0.3--0.4 & 4$^{\circ}$ & 6$^{\circ}$ & 8$^{\circ}$ & 10$^{\circ}$ \\ 
0.4--0.5 & 4$^{\circ}$ & 6$^{\circ}$ & 8$^{\circ}$ & 10$^{\circ}$ \\ 
0.5--0.6 & 2$^{\circ}$ & 3$^{\circ}$ & 4$^{\circ}$ & 5$^{\circ}$ \\ 
0.6--0.7 & 2$^{\circ}$ & 3$^{\circ}$ & 4$^{\circ}$ & 5$^{\circ}$ \\ \hline
\end{tabular}}

\clearpage

\clearpage

\clearpage

\clearpage

\clearpage

\clearpage

\clearpage

\clearpage

\clearpage

\clearpage


\clearpage

\begin{thebibliography}{}

\bibitem{} Baugh, C. M. \& Efstathiou, G. 1993, \mnras, 265, 145

\bibitem{} Blandford, R. D., Saust, A.-B., Brainerd, T. G., \&
Villumsen, J. V. 1991, \mnras, 251, 600

\bibitem{} Brainerd, T. G. 2002, in ``New Cosmological Data and the
Values of the Fundamental Parameters'', proceedings of IAU Symposium 201,
eds.\ A. Lasenby \& A.\ Wilkinson, in press

\bibitem{a1} Brainerd, T.G., Blandford, R.D., \& Smail, I. 1996, ApJ, 
466, 623 (BBS)

\bibitem{} Brainerd, T. G. \& Blandford, R. D. 2002, in ``Dark Matter and
Gravitational Lensing'', eds.\ F.\ Courbin \& D. Minniti,
Springer--Verlag, in press

\bibitem{} Brainerd, T. G. \& Smail, I. 1998, \apj, 494, L137

\bibitem{a2} Brainerd, T.G., \& Wright, C.O. 2000, 
astro-ph/0006281

\bibitem{} Cohen, J. 2002, ApJ, 567, 672

\bibitem{} Cohen, J., Hogg, D. W., Blandford, R., Cowie, L. L., Hu, E.,
Songaila, A., Shopbell, P., \& Richberg, K. 2000, ApJ, 538, 29

\bibitem{a3} Cohen, J., Blandford, R., Hogg, D.W., Pahre, M.A., 
\& Shopbell, P.L. 1999a ApJ, 512, 30

\bibitem{a4} Cohen, J., Hogg, D.W., Pahre, M.A., 
Blandford, R., Shopbell, P.L., \& Richberg, K.  1999b ApJS, 120, 171

\bibitem{a5} Dell'Antonio, I.P. \& Tyson, J.A. 1996, ApJ, 473, L17

\bibitem{} Dubinski, J. \& Carlberg, R. 1991, \apj, 378, 496

\bibitem{a6} Ebbels, T. 1998, PhD Thesis, University of Cambridge

\bibitem{} Fabricant, D. \& Gorenstein, P. 1983, \apj, 267, 535

\bibitem{a7} Fich, M. \& Tremaine, S. 1991, ARA\&A, 29, 409

\bibitem{a8} Fischer, P., McKay, T.A., Sheldon, E.,
Connolly, A., Stebbins, A.,
Frieman, J.A., 
Jain, B., Joffre, M.,Johnston, D., Bernstein, G., Annis, J.,Bahcall, N.A., 
Brinkmann, J., Carr, M.A.,Csabai, I., Gunn, J.E., Hennessy, G.S.,Hindsley, 
R.B., Hull, C., Ivezic, Z.,Knapp, G. R., Limmongkol, S.,Lupton, R.H., 
Munn, J.A., Nash, T.,Newberg, H.J., Owen, R., Pier, J.R.,Rockosi, C.M., 
Schneider, D.P.,Smith, J.A., Stoughton, C.,Szalay, A.S., Szokoly, G.P.,
Thakar, A.R., Vogeley, M.S.,Waddell, P., Weinberg, D.H.,York, D.G., 
The SDSS Collaboration, 2000, AJ, 120, 1198

\bibitem{} Geiger, B. \& Schneider, P. 1999, \mnras, 302, 118

\bibitem{a9} Griffiths, R.E., Casertano, S., Im, M., \& Ratnatunga, K.U.
1996, MNRAS,  282, 1159

\bibitem{a10} Hoekstra, H. 2000, PhD Thesis, University of Groningen

\bibitem{} Hoekstra, H., Yee, H. K. C. \& Gladders, M. D. 2002,
astro-ph/0205205

\bibitem{a11} Hogg, D.W., Cohen, J.G., Blandford, R., Gwyn, S.D.J., 
Hartwick, F.D.A., Mobasher, B., 
Mazzei, P., Sawicki, M., Lin, H.,Yee, H.K.C., Connolly, A.J., Brunner, R.J., 
Csabai, I., Dickinson, M., SubbaRao, M.U., Szalay, A.S., Fernandez-Soto, A., 
Lanzetta, K.M., \& Yahil, A. 1998, AJ, 115, 1418

\bibitem{} Hogg, D. W., Pahre, M. A., Adelberger, K. L., Blandford, R.,
Cohen, J. G., Gautier, T. N., Jarrett, T., Neugebauer, G., \& Steidel,
C. C. 2000, ApJS, 127, 1

\bibitem{a12} Hudson, M.J., Gwyn, S. D. J., 
Dahle, H. \& Kaiser, N. 1998, ApJ, 
503, 531

\bibitem{} Infante, L. \& Pritchett, C. J. 1995, \apj, 439, 565

\bibitem{} Jaunsen, A. O. 2000, PhD thesis, University of Oslo

\bibitem{} Kochanek, C. S. 2001, astro-ph/0106495

\bibitem{} Keeton, C. R., Kochanek, C. S. \& Falco, E. E. 1998, \apj,
509, 561

\bibitem{a14} Kormann, R., Schneider, P., \& Bartelmann, M.1994, A \& A, 
284, 285

\bibitem{a15} LeF$\mathrm{\grave{e}}$vre, O., Hudon, D., Lilly, S.J., 
Crampton, D., Hammer, F. \& Tresse, L. 1996, ApJ, 461, 534

\bibitem{} Maller, A. H., Simard, L., Guhathakurta, P., Hjorth, J., Jaunsen,
A. O., Flores, R. A., \& Primack, J. R. 2000, \apj, 533, 194

\bibitem{} McKay, T. A., Sheldon, E. S., Racusin, J., Fischer, P., Seljak,
U., Stebbins, A., Johnston, D., Frieman, J. A., Bahcall, N., Brinkmann, J.,
Csabai, I., Fukugita, M., Hennessy, G. S., Ivezic, Z., Lamb, D. Q., 
Loveday, J., Lupton, R. H., Munn, J. A., Nichol, R. C., Pier, J. R., \& York,
D. G. 2001, astro-ph/0108013

\bibitem{} McKay, T. A., Sheldon, E. S., Johnston, D., Grebel, E. K., Prada,
F., Rix, H.-W., Bahcall, N. A., Brinkmann, J., Csabai, I., Fukugita, M., 
Lamb, D. Q., \& York, D. G. 2002, ApJ, 571, L85 

\bibitem{} Natarajan, P., Kneib, J.-P. \& Smail, I. 1998, \apj, 499, 600

\bibitem{} Natarajan, P., Kneib, J.-P. \& Smail, I. 2001, in ``Gravitational
Lensing: Recent Progress and Future Goals'', ASP Conf.\ Series 237, eds.
T. G. Brainerd \& C. S. Kochanek, 391

\bibitem{} Natarajan, P. \& Refregier, A. 2000, \apj, 538, L113

\bibitem{a16} Sackett, P.D. 1999, in ``Galaxy Dynamics'', ASP Conf.\
Series 182, eds.\ D. R. Merritt, M.\ Valluri \& J. A. Sellwood, 393

\bibitem{} Schneider, P., Ehlers, J. \& Falco, E. E. 1992, ``Gravitational
Lensing'', (Berlin: Springer--Verlag)

\bibitem{a19} Smail I., Hogg, D.W., Yan, L. \& Cohen, J.G. 1995, ApJ, 449, L105

\bibitem{a20} Smith, D.R., Bernstein, G.M., Fischer, P. \& Jarvis, M. 2001, 
ApJ, 551, 643 

\bibitem{} Stewart, G. C., Fabian, A. C., Nulsen, P. E. J., \& Canizares,
C. R. 1984, \apj, 278, 536

\bibitem{} Villumsen, J. V., Freudling, W., \& da Costa, L. N. 1997,
\apj, 481, 578

\bibitem{} Warren, M. S., Quinn, P. J., Salmon, J. K., \& Zurek, W. H.
1992, \apj, 399, 405

\bibitem{a23} Wilson, G., Kaiser, N., Luppino, G.A., \& Cowie, L.L.
2001, ApJ, 555, 572

\bibitem{} Williams, R. E., Blacker, B., Dickinson, M., Dixon, W. V. D., 
Ferguson, H. C., Fruchter, A. S., Giavalisco, M., Gilliland, R. L., Heyer, I.,
Katsanis, R., Levay, Z., Lucas, R., McElroy, D. B., Petro, L., Postman, M.,
Adorf, H.-M., \& Hook, R. 1996, AJ, 112, 1335

\bibitem{a24} Zaritsky, D. \& White, S.D.M. 1994, ApJ, 435, 599

\bibitem{a25} Zaritsky, D., Smith, R., Frenk, C., \& White, S.D.M. 1997, ApJ, 
478, 39

\end{thebibliography}
\end{document}